# Localization of Dummy Data Injection Attacks in Power Systems Considering Incomplete Topological Information: A Spatio-Temporal Graph Wavelet Convolutional Neural Network Approach

**Zhaoyang Qu[1,5], Yunchang Dong[2*], Yang Li[1,6*], Siqi Song[3], Tao Jiang[1], Min Li[1], Qiming Wang[4], Lei Wang[4], Xiaoyong Bo[1], Jiye Zang[4], Qi Xu[4]**

1 School of Electrical Engineering, Northeast Electric Power University, Jilin, China
2 Jilin Electric Power Research Institute of State Grid, Changchun, China
3 Northeast Electric Power Design and Institute Co., Ltd. of China Power Engineering Consulting Group
4 School of Computer, Northeast Electric Power University, Jilin, China
5 Jilin Engineering Technology Research Center of Intelligent Electric Power Big Data Processing, Jilin, China
6 Department of Civil and Environmental Engineering, The Hong Kong University of Science and Technology, Hong Kong, China

* Correspondence: Yunchang Dong, e-mail: 1202000002@neepu.edu.cn; Yang Li, email: liyang@neepu.edu.cn.

This work was supported by the National Natural Science Foundation of China (No. 52377081) and Natural Science Foundation of Jilin Province (20220101234JC).

**ABSTRACT** The emergence of novel the dummy data injection attack (DDIA) poses a severe threat to the secure and stable operation of power systems. These attacks are particularly perilous due to the minimal Euclidean spatial separation between the injected malicious data and legitimate data, rendering their precise detection challenging using conventional distance-based methods. Furthermore, existing research predominantly focuses on various machine learning techniques, often analyzing the temporal data sequences post-attack or relying solely on Euclidean spatial characteristics. Unfortunately, this approach tends to overlook the inherent topological correlations within the non-Euclidean spatial attributes of power grid data, consequently leading to diminished accuracy in attack localization. To address this issue, this study takes a comprehensive approach. Initially, it examines the underlying principles of these new DDIAs on power systems. Here, an intricate mathematical model of the DDIA is designed, accounting for incomplete topological knowledge and alternating current (AC) state estimation from an attacker's perspective. This model aims to mitigate the vulnerabilities associated with direct current (DC) flow-based attack data generation methods that can be readily detected. Subsequently, by integrating a priori knowledge of grid topology and considering the temporal correlations within measurement data and the topology-dependent attributes of the power grid, this study introduces temporal and spatial attention matrices. These matrices adaptively capture the spatio-temporal correlations within the attacks. Leveraging gated stacked causal convolution and graph wavelet sparse convolution, the study jointly extracts spatio-temporal DDIA features. This methodology significantly enhances the dynamic correlation mining capability while improving computational efficiency in the neighborhood. Finally, the research proposes a DDIA localization method based on spatio-temporal graph neural networks, allowing for adaptation to changing power system topologies. The accuracy and effectiveness of the DDIA model are rigorously demonstrated through comprehensive analytical cases. It is unequivocally verified that the proposed localization method rapidly and effectively detects and locates DDIA while exhibiting superior accuracy, robustness, and generalization capabilities.

**INDEX TERMS** power system attack positioning; dummy data injection attack; spatio-temporal graph neural network; graph wavelet convolution; gated causal convolution

## 1. Introduction

The integration of advanced information and communication technologies in power systems has spurred the intelligent development of new power systems, giving rise to tightly coupled cyber-physical systems [1-3]. Malicious actors exploit vulnerabilities in sensors, communication infrastructure, and smart metering equipment to launch pre-designed attacks, increasing the cybersecurity risks faced by power systems. Among these threats, the false data injection attack (FDIA) is particularly concerning [4]. Recent research has revealed the emergence of an evolved attack, the DDIA [5], introducing novel challenges to power system security.

The enhanced DDIA conceals injected false data within the regular measurement data, distinguishing it from the earlier FDIA, where injected data was flagged as anomalies. Consequently, the injected false data are now more inconspicuous, blending closely with genuine measurements. Employing conventional clustering and distance-based attack detection and localization methods becomes ineffective because these fake data no longer stand out. To mitigate the adverse impacts of such attacks on the secure and stable operation of the power grid, timely detection of the attack's origin, isolation of the compromised bus, and reconfiguration of system currents become critical for sustaining the power system's reliability and economic dispatch.

The DDIA was initially proposed by Liu et al. [8], highlighting the challenge of concealing malicious data within regular measurement data, thus amplifying the complexity of detecting erroneous data. In response to this attack, reference [9] ranked the importance of load nodes based on active power and topological connectivity metrics, establishing robust protection measures that make it challenging to craft effective dummy data, providing an optimal defense against DDIA. However, for the DDIA occurring within the system, the existing research has yet to produce a more effective attack detection and localization method, with the primary focus being on FDIA defense.

Regarding attack modeling, Liu et al. introduced the concept of FDIA and dissected its attack mechanism [10-12]. Subsequent works delved into the attack principles and conditions of the FDIA, forming the foundation for subsequent strategies related to detection, localization, and defense. Reference [13] constructed an AC-based FDIA model encompassing complete and incomplete topology information. Reference [14] proposed an AC-based FDIA generation method that didn't rely on network topology information, instead utilizing adversarial generative networks. These studies demonstrated that the constructed attacks could effectively evade bad data detection under varying measurement error conditions.

From the perspective of power system regulators, accurate and efficient detection of the FDIA is of utmost importance when countering such threats. Detection methods encompass a variety of approaches, including the use of the Kalman filter and its enhancements, low-rank decomposition methods, hybrid Gaussian distribution techniques, Euclidean detectors, Kullback-Leibler distance calculations, sparse optimization, similarity matching, data mining, machine learning, and deep learning [15-18].

In practice, determining the location of an attack is essential for devising effective defensive countermeasures. However, compared to attack detection, relatively little research has been dedicated to attack localization. Reference [19] addresses the FDIA detection problem by employing a model-free deep learning approach that combines the Convolutional Neural Network (CNN) and traditional bad data detectors for accurate attack localization under different noise and attack conditions. Reference [20] proposes an FDIA localization method based on the extreme learning machine but is limited by the number of implied layers, making it challenging to handle high-dimensional and voluminous power system measurement data. Reference [21] introduces a multi-grain FDIA localization algorithm based on graph theory techniques, while reference [22] creates a distributed interval observer for each measurement device to locate attacks through a logical discriminant matrix. Reference [23] presents an effective FDIA location identification strategy based on a CNN-BiLSTM model as a multi-label classification method, demonstrating sensitivity and robustness. Reference [24] employs an improved capsule network to detect and locate FDIAs, maintaining stability and accuracy in attack location detection under varying attack sparsity and interference size. Reference [25] proposes a cluster-based architecture for the detection and location identification of false data injection attacks on smart grids. It represents grid state vectors as multivariate time series and uses vector autoregression for the FDIA detection and localization. To address the challenge of multimodal probability distributions of measured values and state variables in state estimation for the FDIA localization, reference [26] constructs an auto-encoder-based generative adversarial network to generate normal distributions of multimodal measurements offline, thus creating candidate sets for locating and recovering falsified measurements.

In summary, all of the aforementioned studies focus on FDIA localization, lacking similar research on the new type of DDIA. This lack of research leads to a deficiency in scalability and automated intelligence for DDIA localization models. The robustness and adaptability of localization models for DDIA are yet to be demonstrated, and deploying detectors throughout the system can only determine the regional location of the attack while wasting deployment resources. The spatio-temporal evolution of attacks is characterized by complexity and cross-domain interactions. Attack trends are strongly correlated with timing and grid topology, evolving over time and space, potentially leading to cascading failures in both information and physical domains. Thus, discerning the intent behind attack behaviors is challenging. Existing research on localization methods for the FDIA often overlooks the non-Euclidean spatial correlation characteristics of grid topology associations, resulting in imprecise attack location determination. Additionally, meter readings in the grid measurement space are correlated due to the topology and distribution of power measurement devices, making it inaccurate to assume an independent and identical distribution of meter readings for data-driven models.

To address these research gaps, this paper proposes a DDIA localization method based on spatio-temporal graph convolutional neural networks. To the best of the authors' knowledge, this is the first paper to explore DDIA localization methods for power systems. The main contributions of this paper are as follows:

(1) In order to establish an accurate mathematical model for the DDIA in power systems, a novel DDIA model is developed, taking into account incomplete topology information in the context of AC state estimation. Through a comprehensive analysis of DDIA's attack principles, which involve concealing itself within normal measurements and evading bad data detection, this model adheres to consistency and virtuality constraints. It considers more realistic AC flow calculations and introduces the condition that the attacker possesses only partial topology information about a portion of the power grid. This additional condition aligns the attack model more closely with real operational scenarios and conditions.

(2) To fully exploit the temporal and topological spatial correlations within power system DDIA data, a DDIA feature extraction model is introduced. This model combines the spatio-temporal attention mechanism with a graph wavelet convolutional neural network. The spatio-temporal attention mechanism adaptively captures the correlations among attack features and enhances the model's interpretability by incorporating knowledge of the power grid's topology into the data-driven approach. It takes into consideration the power system's topological connections and the temporal flow correlation of non-Euclidean spatial distances in measurement data, allowing for a comprehensive exploration of the spatial and temporal evolution features of the attack. This, in turn, aids in accurately pinpointing the location of the attack.

(3) A frequency domain transformation of attack features is implemented using a graph wavelet neural network, enhancing computational efficiency, accuracy, and flexibility. The scale factor is reduced through graph wavelet transformation to enable a more flexible feature range analysis.

(4) To achieve precise localization of DDIAs while ensuring the method's robustness and adaptability, a multi-tag DDIA end-to-end localization method with a sandwich structure is designed. This method can detect the presence of DDIA and identify the specific location of DDIA occurrence. Experimental analysis of the algorithm demonstrates the method's effectiveness in both detecting and locating DDIAs, showcasing high localization performance, and its ability to adapt to noise interference, as well as variations in scale and topology.

## 2. The DDIA modeling of power systems considering AC flow and incomplete topology information

The DDIA represents an exceptionally covert form of attack, characterized by its distinctive attack principles, which must adhere to the consistency and virtualization constraints of measurement data. To precisely replicate real-world attack scenarios, we incorporate the complete AC flow equation, instead of relying on the DC approximation model. Furthermore, recognizing the constraints imposed by limited attack resources, our mathematical model for the DDIA accounts for incomplete topology information, enabling the attack to bypass the bad data detector.

### *2.1. AC status estimation and bad data detection*

Power system operation generates measurement data, including bus voltage amplitudes, power injection, and line power. However, these measurements are susceptible to errors arising from factors such as equipment inaccuracies, communication link noise, and human-induced interference [28]. The DC flow model simplifies various conditions and may not adequately meet the safety requirements of real-world operations. Consequently, this paper centers its attention on the DDIA in the context of AC state estimation. In a power system with multiple buses and transmission lines, the relationship between measurements and state variables, as determined by the AC state estimation, can be expressed as follows:

$$z = h(x) + e \tag{1}$$

where: $z$ represents the collected measurement vector, encompassing parameters such as bus active power, bus reactive power, branch active flow, branch reactive flow, and more. $x$ denotes the estimated state vector, encompassing voltage magnitude and voltage phase angles. The reference bus is conventionally set as the reference point (0). $h(\cdot)$ represents the measurement equation with nonlinear dependencies, which is jointly influenced by the network's topology, connectivity relationships, and transmission line parameters, as depicted in equation (2). $e$ signifies the measurement error, typically manifesting as additive noise with a covariance of $R$ .

$$h(x) = \begin{bmatrix} P_{ij}\left(\theta_{ij}, V_{ij}\right) \\ Q_{ij}\left(\theta_{ij}, V_{ij}\right) \\ P_i\left(\theta_{ij}, V_{ij}\right) \\ Q_i\left(\theta_{ij}, V_{ij}\right) \\ V_i\left(V_i\right) \end{bmatrix} \tag{2}$$

The relationship governing the flow equation between measurement and state variables is as follows:

$$P_i = \sum_{j=1}^{n} V_i V_j \left(G_{ij}\cos\left(\theta_{ij}\right) + B_{ij}\sin\left(\theta_{ij}\right)\right) \tag{3}$$

$$Q_i = \sum_{j=1}^{n} V_i V_j \left(G_{ij}\sin\left(\theta_{ij}\right) + B_{ij}\cos\left(\theta_{ij}\right)\right) \tag{4}$$

$$p_{ij} = -V_i^2 G_{ij} + V_i V_j \left(G_{ij}\cos\left(\theta_{ij}\right) + B_{ij}\sin\left(\theta_{ij}\right)\right) \tag{5}$$

$$q_{ij} = V_i^2 B_{ij} + V_i V_j \left(G_{ij}\sin\left(\theta_{ij}\right) - B_{ij}\cos\left(\theta_{ij}\right)\right) \tag{6}$$

Where: $P_i$ represents the active power injection at bus $i$ . $Q_i$ represents the reactive power injection at bus $i$ . $p_{ij}$ represents the active power flow from bus $i$ to bus $j$ . $q_{ij}$ represents the reactive power flow from bus $i$ to bus $j$ . $V_i$ is the voltage magnitude at bus $i$ . $V_j$ is the voltage magnitude at bus $j$ . $G_{ij}$ and $B_{ij}$ are the real and imaginary parts of the derivative matrix. $\theta_{ij}$ is the phase angle difference between buses $i$ and $j$ .

With known values of $z$ , $h(\cdot)$ , and $R$ , this equation can be solved using the most comprehensive weighted least squares method to estimate the system's state variables.

$$\hat{x} = \arg\min_{x} [z - h(\hat{x})]^T R^{-1} [z - h(x)] \tag{7}$$

Where: $R$ represents the diagonal matrix of measurement noise covariances. $\hat{x}$ stands for the estimated state vector. Equation (7) can be solved iteratively using the Newton-Raphson method.

One common technique employed within this method is the maximum normalized residual test. The residuals are defined as follows [29]:

$$\| r \|_2 = \left\| z - h(\hat{x}) \right\|_2 \tag{8}$$

If the disparity between the observed measurement $z$ and the measurement $z'$ inferred from the state estimate is within a specified threshold value, the measurement is considered plausible. Conversely, if the difference exceeds this threshold, the measurement is classified as bad data.

The necessity of using the proposed method to detect the FDIA is as follows.

1) Nature of FDIA: Firstly, it is crucial to clarify that FDIAs are sophisticated attacks designed to bypass conventional bad-data-detection systems. Unlike random errors or faults, FDIA strategically manipulates data to align with the physical laws governing power systems, like power flow equations. This characteristic enables the FDIA to mimic legitimate operational data, thereby eluding detection by standard state estimation procedures.

2) Limitations of Conventional Detection: Conventional state estimation methods rely heavily on detecting anomalies that deviate significantly from expected operational patterns. FDIA, however, exploits this by ensuring that the injected false data do not present as anomalies. Therefore, while these systems are effective against random faults or errors, they are not equipped to identify strategically manipulated data that adheres to the system’s operational norms.

3) Advantages of Proposed Method: Our proposed method utilizes a spatio-temporal graph wavelet convolutional neural network approach. This method is specifically designed to detect the subtle manipulations characteristic of the FDIA. It captures the intricate spatial and temporal correlations within the power system data, which are often overlooked by traditional detection methods. This heightened sensitivity allows for the identification of FDIAs that are crafted to be indistinguishable from normal operational data.

4) Justification for Complexity: The complexity of our method is justified by its enhanced capability to identify these covert attacks. The conventional state estimation methods, while effective in many scenarios, fall short in detecting FDIAs due to their design to blend seamlessly with legitimate data. Our method provides an additional layer of security, helping to safeguard power systems against these advanced forms of cyber-attacks.

In conclusion, the complexity of our proposed method is a necessary response to the evolving sophistication of FDIAs, which are designed to circumvent traditional detection mechanisms. We believe that our approach significantly contributes to improving the resilience of power systems against such covert threats.

### *2.2. The DDIA principle*

Traditional FDIA involves constructing an attack vector that complies with the flow constraints and superimposing it onto normal measurement data without significantly altering the measurement residuals. This approach effectively evades bad data detection.

In contrast, the DDIA introduced in this paper is an extension of FDIA. It entails crafting the attack vector in such a way that the tampered measurement data closely resembles the normal measurement data in terms of spatial characteristics. In other words, from a spatial perspective, the attacked data appears extremely similar to the normal measurements, making it challenging to identify and isolate the abnormal data. This characteristic is referred to as "dummy datas."

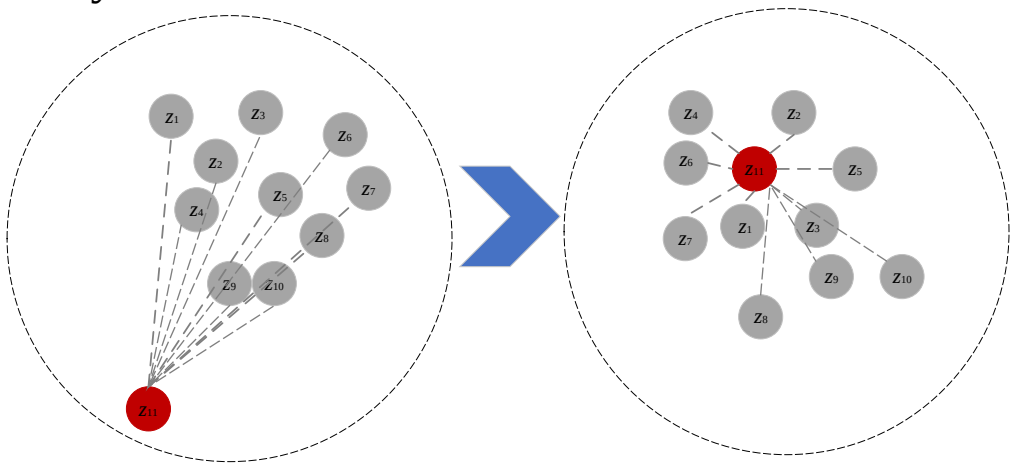

(a) FDIA (b) DDIA

Fig. 1. Difference between FDIA and DDIA

As illustrated in Fig. 1(a), the data injected by FDIA stands as an outlier, exhibiting a substantial spatial separation from other normal data in terms of Euclidean spatial distance. Consequently, existing clustering and machine learning methods can readily identify the presence of the FDIA. However, in Fig. 1(b), the DDIA-injected data maintains a smaller distance from other normal data, allowing the attacked measurements to blend in with the normal data. As a result, existing clustering and machine learning methods struggle to distinguish the DDIA from the normal data.

Hence, both the DDIA and the FDIA must adhere to the same physical laws' consistency, ensuring that the tampered measurements continue to satisfy Kirchhoff's voltage law and Kirchhoff's current law. It's imperative that the total residuals in the AC state estimation remain unchanged.

Furthermore, the DDIA introduces an additional condition known as spatial distance virtuality. This means that the tampered measurement data, when mixed with normal measurement data, should not be identified as outliers. This constraint involves maintaining a minimum spatial distance between the injected data and other normal data.

In addition, DDIA should have an impact on the safe operation of the power system. The injection of such virtual data alone should not meet the attacker's objectives; it should result in adverse consequences, such as power system line overloads and other unsafe operations.

Moreover, the DDIA limited attack area sets it apart from traditional FDIAs. The attacker possesses restricted information about the network's topology and line parameters, and the attack area is confined. The injected attack vectors should not cause changes in the system state and measurements outside the attack area. In other words, the alterations in the system's currents resulting from the attack should remain confined within the specified attack area.

### *2.3. The DDIA model*

Not all types of cyberattacks can be executed successfully, and many critical measurement points are equipped with robust defenses. For instance, power plant system control centers often maintain real-time communication, making it exceptionally challenging to manipulate generator output power values [30]. Therefore, the attack scenario described in this paper assumes that only the power injection at load buses can be targeted, while the measurements at zero-injected buses remain untouched. The attacker's objective is to maintain the total net power injection measurements within the attack area unchanged, thereby keeping the total load constant. To achieve this, the attacker artificially increases the power injections at certain buses while decreasing them at others, aligning with actual operating conditions.

The concealment conditions for the attack necessitate that line flows, voltage magnitudes at boundary nodes, and voltage phase angles at boundary buses remain constant before and after the attack. Consequently, the load redistribution constraint for the DDIA can be formulated as follows:

$$\min \sum_{a \in A} c_a V_a \tag{9}$$

$$\begin{bmatrix} S'_C \\ V'_B \\ \Delta\theta'_B \end{bmatrix} = \begin{bmatrix} S_C \\ V_B \\ \Delta\theta_B \end{bmatrix} \tag{10}$$

$$\sum_{a \in A} \Delta D_a = 0 \tag{11}$$

$$-\delta S_a^l \le \Delta D_a \le \delta S_a^l \tag{12}$$

Where: $A, B, C$ represent the sets of nodes in the attack area, boundary nodes, and contact lines, respectively. a is the targeted node within the attack area $A$. $c_a$ denotes the non-negative weight coefficient assigned to region $A$. $S_C$ and $S'_C$ are the power flows of the contact line before and after the attack, respectively. $V_B$ and $V'_B$ are the voltage amplitudes of boundary nodes before and after the attack. $\Delta\theta_B$ and $\Delta\theta'_B$ are the voltage phase angle difference of the boundary node. $D_a$ represents the attack magnitude, which signifies the load variation per bus in the attack area. $\delta$ is the weight threshold allowed for the original load change. $S_a^l$ represents the original load per bus that remains unaffected by the attack.

The DDIA is constrained by the AC flow, incorporating topological information and line parameters specific to the attack area, in conjunction with boundary information pertaining to the non-attack area.

$$S^l = P^l + iQ^l \tag{13}$$

$$g_P\left(\theta, V, P^l\right) = 0 \tag{14}$$

$$g_Q\left(\theta, V, Q^l\right) = 0 \tag{15}$$

$$h_f(\theta, V) \le 0 \tag{16}$$

$$h_t(\theta, V) \le 0 \tag{17}$$

Where: $\theta$ represents the voltage phase angle. $V$ represents the voltage magnitude. $p^l$ signifies active power. $Q^l$ represents reactive power. These variables are decision variables in the context of the model. $g_P$ denotes the nonlinear constraint for active power at the node. $g_Q$ represents the nonlinear equation constraint for reactive power balance at the node. $h_f$ and $h_t$ are the flow nonlinear constraints from one node to another node.The attack vectors generated by the attack model adhere to Kirchhoff's current and voltage laws, ensuring consistency.

To guarantee that the tampered measurement data satisfy the "virtualness" constraint, normal measurement data matrices are generated using the Monte Carlo simulation method. The tampered measurement value $z$ is then constructed by identifying a point that closely matches the normal measurements. The DDIA is solved to minimize the total distance between the attacked sample and each normal measurement sample [5], [8].

$$\min \left\|z' - z_1\right\|_2 + \left\|z' - z_2\right\|_2 + \ldots + \left\|z' - z_n\right\|_2 \tag{18}$$

To address the equation mentioned above, the nonlinear problem presented in (18) is reconfigured as follows:

$$\min \left|z' - z_1\right| + \left|z' - z_2\right| + \ldots + \left|z' - z_n\right| \tag{19}$$

This nonlinear problem is converted into a linear programming problem by introducing additional auxiliary variables $\xi_1$, $\xi_2$, ..., $\xi_n$, as follows:

$$\min 1^T \xi_1 + 1^T \xi_2 + \ldots + 1^T \xi_n \tag{20}$$

Bound by (18)-(20):

$$\begin{cases} \xi_1 \ge z' - z_1 & \xi_1 \ge -\left(z' - z_1\right) \\ \xi_2 \ge z' - z_2 & \xi_2 \ge -\left(z' - z_2\right) \\ & \vdots \\ \xi_n \ge z' - z_n & \xi_n \ge -\left(z' - z_n\right) \end{cases} \tag{21}$$

where: $1^T$ represents a row vector comprising all its elements. By computing the dot product, all the elements in $\xi$ are summed together.

The introduction of dummy data can potentially result in line flow overloads:

$$\Delta F^l / F^l_{\max} \ge \sigma \tag{22}$$

where: $\Delta F^l$ represents the change in line flow. $F^l_{\max}$ is the maximum line flow. $\sigma$ denotes the overload level.

Hence, the optimal injection problem for dummy data can be expressed as follows:

$$\text{LP1} \ \min 1^T \xi_1 + 1^T \xi_2 + \cdots + 1^T \xi_n \tag{23}$$

The DDIA can be covertly integrated into the normal measurements while adhering to the specified constraints. This process yields a comprehensive dataset of quantitative measurements, encompassing the attacked data. The attack modeling is reformulated as an optimization problem, and the optimal solution represents an attack sample. In this reimagined attack modeling, akin to an implicit function, the approach no longer relies on $a = HC$ to construct the DDIA.

It's crucial to highlight that in the spatial dimension, the injected dummy data closely resembles normal measurements. However, a significant distinction lies in the potential consequences, including exceeding flow limits. Moreover, the magnitude of the attack can directly influence the extent of the resulting harm. Consequently, many existing distance-based data mining algorithms and machine learning algorithms may prove ineffective. Hence, it's imperative to develop localization methods tailored to address the unique challenges posed by the DDIA.

## 3. The DDIA spatio-temporal feature extraction based on attention mechanism and spatio-temporal graph wavelet convolution

The operation of the power system, post-DDIA attack, results in measurement data that incorporates not only temporal flow characteristics but also non-Euclidean spatial dependencies stemming from topological connections. This modeling approach treats the power system as a graph neural network model.

### *3.1. Graph network model of power system*

Power systems exhibit a graph-like structure that can be represented as graph network models comprising vertices and edges. Traditional deep learning models like CNN and Recurrent Neural Networks (RNN) are not suitable for processing graph data [31]. In the context of a particular power system network, the topological graph is abstracted as a connected undirected weighted graph network model $G$:

$$G = (U, E, W) \tag{24}$$

Where: $U$ is the set of $N$ nodes (buses). $E$ is the set of $F$ branches and transformers, denoting the connections between nodes. $W$ is the weighted adjacency matrix of the graph $G$, typically represented by the admittance matrix. A larger value of $W$ signifies a stronger association between nodes, reflecting prior knowledge about the topological connections within the power system. This association can be adjusted based on specific changes in the topology.

### *3.2. The DDIA spatio-temporal correlation capture*

To automatically capture the spatio-temporal dynamic correlations among nodes, temporal and spatial attention matrices are established. These matrices assign more significant weights to multidimensional measurement feature vectors that require attention. This prioritization allows for the extraction of more valuable spatio-temporal information.

1) *Time Attention Matrix*

In the temporal dimension, measurement data exhibits a typical temporal correlation flow. There exists backward and forward temporal dependencies in the system's operational state across different time segments. To capture this temporal correlation, a temporal attention matrix $A$ is used for the DDIA.

$$A = V_e \times \sigma(((\chi^{(r\text{-}1)})^T U_1) U_2 (U_3 \chi^{(r\text{-}1)}) + b_e) \tag{25}$$

$$A'_{i,j} = \frac{\exp(A_{ij})}{\sum_{j=1}^{T_{r-1}} \exp(A_{ij})} \tag{26}$$

Where: $\mathrm{X}^{(r-1)} = (X_1, X_2, \ldots, X_{T_{r-1}}) \in \mathbb{R}^{N \times C_{r-1} \times T_{r-1}}$ represents the input data of the $r$-th spatio-temporal module. $C_{r-1}$ is the number of channels for the input measurement data of the $r$ -th layer. $T_{r-1}$ represents the length of the $r$ -th layer's time dimension. $V_e, b_e \in \mathbb{R}^{T_{r-1} \times T_{r-1}}$ , $U_1 \in \mathbb{R}^{N}$ , $U_2 \in \mathbb{R}^{C_{r-1} \times N}$ , and $U_3 \in \mathbb{R}^{C_{r-1}}$ are all learnable parameters. $\sigma$ represents the sigmoid activation function. $\sigma$ denotes the strength of temporal correlation between time $i$ and time $j$ .The $E$ values are normalized using the Softmax operation.

*2) Spatial attention matrix*

In the spatial dimension, measurements on different nodes exert mutual influence on the power system state. Similar to the temporal attention mechanism described above, a spatial attention matrix $B$ is used in the DDIA to capture spatial correlations among measurement data.

$$B = V_s \cdot \sigma(((\mathrm{X}^{(r-1)} W_1) W_2 (W_3 \mathrm{X}^{(r-1)})^T + b_s) \quad (27)$$

$$B_{i,j}^{'} = \frac{\exp\left(B_{i,j}\right)}{\sum_{j=1}^{N} \exp\left(B_{i,j}\right)} \quad (28)$$

where: $V_s, b_s \in \mathbb{R}^{N \times N}$ , $W_1 \in \mathbb{R}^{T_{r-1}}$ , $W_2 \in \mathbb{R}^{C_{r-1} \times T_{r-1}}$ , and $W_3 \in \mathbb{R}^{C_{r-1}}$ are learnable parameters. $S_{i,j}^{'}$ denotes the strength of spatial correlation between node $i$ and node $j$ . The Softmax operation is applied to normalize $S$ .When performing the graph convolution operation, the adjacency matrix is multiplied by the spatial attention matrix $B_{i,j}^{'}$ to collectively adjust the magnitude of the influence weights of the nodes. The spatial attention matrix, obtained after normalization, is directly used as input for the graph convolution calculation.

### 3.3. Temporal feature extraction by gated dilated causal convolution

In this paper, the temporal dynamic correlation features of the DDIA are rapidly extracted in the time dimension using the inflated causal convolutional layer, augmented by gating units. In the extended causal convolutional network [32], the perceptual field range can be expanded by increasing the number of network layers, thereby enhancing the ability to process a wider range of temporal data. The inflated causal convolution introduces a dilation rate, which allows the filter kernel to skip parts of the input data, thus enabling it to filter a broader region beyond its immediate vicinity. At a power system graph node, given a time series $X$ and a filter kernel $y$ , the operation for inflated causal convolution is as follows:

$$X *_{ex} y[t] = \sum_{s=0}^{k} y[s] X\left[t - d * s\right] \quad (29)$$

where: $*_{ex}$ represents the inflated causal convolution operation. $k$ is the convolution kernel size. $d$ is the inflation factor, which determines the skip distance in the input. $s$ is the scaling factor, a parameter that governs the neighborhood node range.

The inflated causal convolution can process temporal data in a parallel and non-recursive manner. This effectively mitigates the problem of gradient explosion when stacking network layers to capture correlations in long-time series measurement data. To enhance the efficiency of temporal convolution, a gating unit is employed to capture nonlinear relationships within the measurement data in the time dimension. The temporal convolution layer with the gating unit is defined as follows:

$$Z = \delta\left(\Theta_1 *_{ex} \chi\right) \odot \delta\left(\Theta_2 *_{ex} \chi\right) \quad (30)$$

where: $\delta$ represents the sigmoid activation function. $X$ is the input to the gating unit.

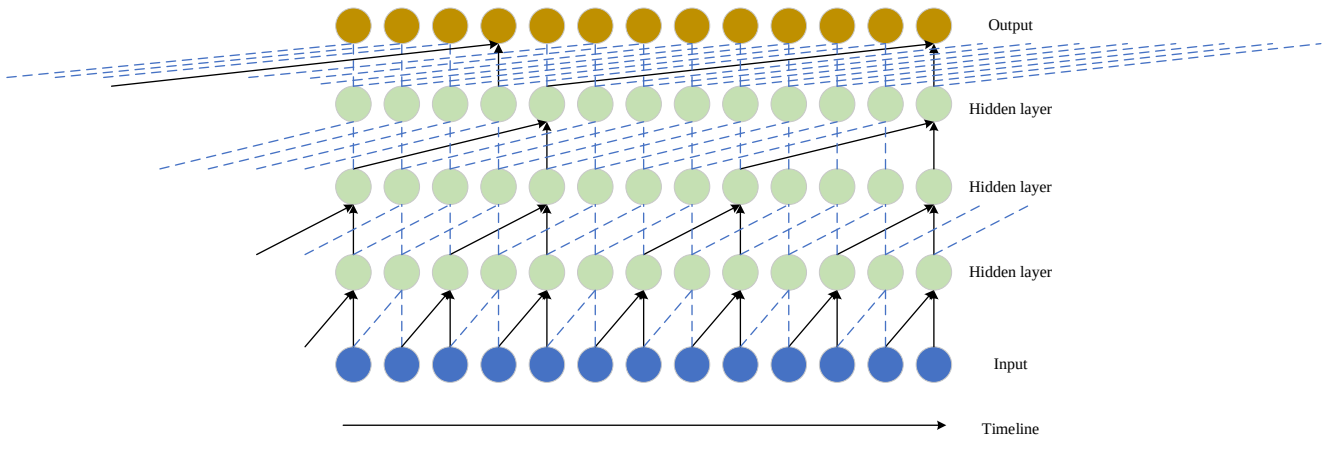

Fig. 2. Inflated causal convolution.

### 3.4. Spatial Feature Extraction by Convolution of Spectral Wavelet Frequency Domain

In the context of power systems, the topological structure of line connection information is considered as typical non-Euclidean spatial data. Traditional deep learning methods based on Euclidean distance metrics often overlook this non-Euclidean property. Modeling power system operation data as long vectors containing various feature quantities can be challenging for effectively handling power system graph data, and it may not adapt well to changes in the system's topology. This, in turn, can impact the effectiveness of DDIA detection and localization.

The decision to perform a frequency domain transform on attack features is motivated by several factors:

1. Computational Efficiency: The graph wavelet transform is computationally efficient because it eliminates the need for feature decomposition of the Laplace matrix.

2. Sparsity: The sparsity of the wavelet transform significantly reduces computational complexity.

3. Local Feature Convolution: The operation of local feature convolution enables precise attack localization feature extraction.

4. Scale Control: The introduction of a scale factor allows for fine-tuning the spatial feature extraction scale in the DDIA. This is achieved by constraining the range of neighborhoods in a discrete shortest-path manner.

In the context of the power system graph $Q$ and its adjacency matrix $W$ , the power system topology is characterized by the normalized Laplace matrix $L$ as follows:

$$L = I_n - D^{-\frac{1}{2}} W D^{-\frac{1}{2}} \quad (31)$$

where: $L$ is a real symmetric, semi-positive definite matrix that can be decomposed into $L = U \wedge U^T$ , where $U = [u_0, \cdots, u_{n-1}] \in \mathbb{R}^{n \times n}$ denotes the matrix's eigenvector, the matrix $L$ that can be decomposed into $n$ orthogonal eigenvectors and $\lambda = diag([\lambda_0, \cdots, \lambda_{n-1}]) \in \mathbb{R}^{n \times n}$ corresponds to the $n$ eigenvalues of $G$ ; $I_n \in \mathbb{R}^{n \times n}$ is the unit matrix; $D_{ij} = \sum_j W_{ij}$ .

Research has been carried out involving the utilization of the graph Fourier transform within the framework of a graph convolutional neural network. This approach involves performing convolution operations in both the frequency and spectral domains on the power system graph. The process includes the following steps:

1. Transforming the graph signal, initially defined in the graph node domain, into the frequency and spectral domain using the graph Fourier transform.

2. Conducting feature extraction through a filtering operation performed in the graph frequency and spectral domain.

3. Reverting the graph signal to the node domain via the Fourier

inverse transform.

The graph convolution operation can be expressed as follows, in accordance with reference [33]:

$$x *_G y = U\left(U^T y\right) \odot \left(U^T x\right) = U g_\theta U^T x \tag{32}$$

where: $x$ is the power graph $G$ signal, $y$ is the graph convolution kernel, $\odot$ is the Hadamard product, $*_G$ is the graph convolution operator, and $g_\theta$ is the learnable graph filter kernel. The computational speed is affected by the large computation of equation (30), unfavorable for real-time attack localization.

Using the Fourier transform to perform graph convolution in power system graphs has its limitations:

1. Computational Complexity: The computation of the eigenvalue data and eigenvector matrix obtained through the eigendecomposition of the Laplace matrix has a complexity of $O(n^3)$ , making it time-consuming, especially for large-scale graph computations.

2. Multiplicative Operation: The Fourier transform involves a multiplicative operation between the dense matrix $U$ and the graph signal $x$ , which can result in low operational efficiency.

3. Lack of Localized Features: The graph convolution operation in the Fourier transform considers the entire node domain of the graph, neglecting the localized neighborhood features of a node. This can lead to inadequate attack feature extraction.

These limitations highlight the need for more efficient and localized methods for feature extraction in power system graph analysis.

To address the aforementioned limitations and decrease computational complexity, reference [34] suggested parameterizing $g_\theta$ as a Chebyshev polynomial function and achieving localized convolution through recursive computation.

$$g_\theta *_G x = g_\theta(L)x = \sum_{k=0}^{K-1} \theta_k T_k(L)x \tag{33}$$

where: $\theta \in R^K$ is the Chebyshev approximation polynomial coefficient vector; $K$ is the Chebyshev polynomial order; $T_k(\cdot)$ is the $k$ th order Chebyshev polynomial; and $L$ is the normalized Laplace matrix.

For this equivalent convolutional filter kernel, the larger the $K$ order truncation of the Chebyshev polynomial, the more difficult it is to extract the local neighborhood features of the convolutional DDIA. When $K$ is smaller, it is difficult to approximate the diagonal matrix $g_\theta$ with $n$ free parameters, and the approximation error is larger and affects the feature extraction effect.

To address this issue, this paper suggests using the graph sparse wavelet frequency domain transform to map the power system graph signal (measurement data) from the vertex domain to the frequency domain. This approach relies on graph wavelet transform [35] instead of employing the graph Fourier transformation as in equation (30) to perform the graph convolution operation:

$$x *_G y = \psi_s\left(\psi_s^{-1} y\right) \odot \left(\psi_s^{-1} x\right) = \psi_s g_\theta \psi_s^{-1} x \tag{34}$$

where: $\psi_s = U G_s U^T = (\psi_{s1}, \psi_{s2}, \ldots, \psi_{sn})$ represents the wavelet bases used for the transformation. $G_s = \mathrm{diag}(g(s\lambda_1), g(s\lambda_2), \cdots, g(s\lambda_n))$ is a normalized sparse diagonal array. $g(s\lambda_n) = e^{\lambda_i s}$ refers to the control heat kernel function scale range.

The power system graph wavelet transform $\chi$ , and the graph wavelet inverse transform $x$ can be expressed as:

$$\chi = \psi_s^{-1} x \tag{35}$$

$$x = \psi_s \chi \tag{36}$$

Where $\psi_s^{-1}$ and $\psi_s$ are sparse and sparse matrix operations can be performed, $\psi_s^{-1}$ and $\psi_s$ can be solved by SGWT algorithm for fast polynomial. The complexity of the operation is reduced to $O(mK)$ , while the graph wavelet corresponds to the graph signal of each node, and the mining of the DDIA local features can be achieved through the flexible control of the above $s$ parameters.

The spatially adaptive feature extraction of the DDIA can be accomplished through the aforementioned graph wavelet convolution operation, offering several notable advantages:

1. Efficient Graph Wavelet Transform: The graph wavelet transform eliminates the need for Laplace matrix feature decomposition, resulting in computational efficiency.

2. Enhanced Operational Efficiency: The sparsity of the wavelet transform substantially reduces operational complexity.

3. Improved Localization Accuracy: Local feature convolution operations enhance the precision of attack localization feature extraction.

4. Flexible Spatial Feature Extraction: The DDIA's spatial feature extraction scale can be flexibly controlled by constraining the neighborhood range using a discrete shortest path approach.

The graph wavelet neural network is a multi-layer convolutional neural network. Building upon the graph wavelet convolution calculation method described above, we have developed the graph wavelet convolution layer for spatial feature extraction in the context of the DDIA . For a measurement data tensor $X$ with an input dimension of $n \times p$ , the output at the $m$ th layer is as follows:

$$X_{[:,j]}^{m+1} = \mathrm{ReLU}\left(\psi_s \sum_{i=1}^{p} F_{i,j}^m \psi_s^{-1} X_{[:,i]}^m\right) \tag{37}$$

Where: $X_{[:,j]}^m$ is $X^m$ the $j$ th column; $F_{i,j}^m$ is the diagonal filter matrix learned in the spectral domain, and ReLU is the nonlinear activation function.

Compared to existing model-based and data-based approaches, the attack feature frequency domain transform combines the strengths of both by incorporating grid topology knowledge into attack characterization. Instead of relying solely on data-driven feature learning, this approach enhances the interpretability of data mining, fully captures the spatial topology information of the transmission grid, and can identify hidden spatial distances in the case of DDIA through the frequency domain transform. The spectral frequency domain transform maintains the computational efficiency of machine learning while reducing computational complexity compared to traditional modeling methods. This approach, which takes into account the topology calculation method, is also adaptable for detecting and localizing DDIAs in various grid topologies, even when the grid structure and size change. It can reflect changes in the physical dynamics of the grid through iterative calculations that incorporate topological structure information.

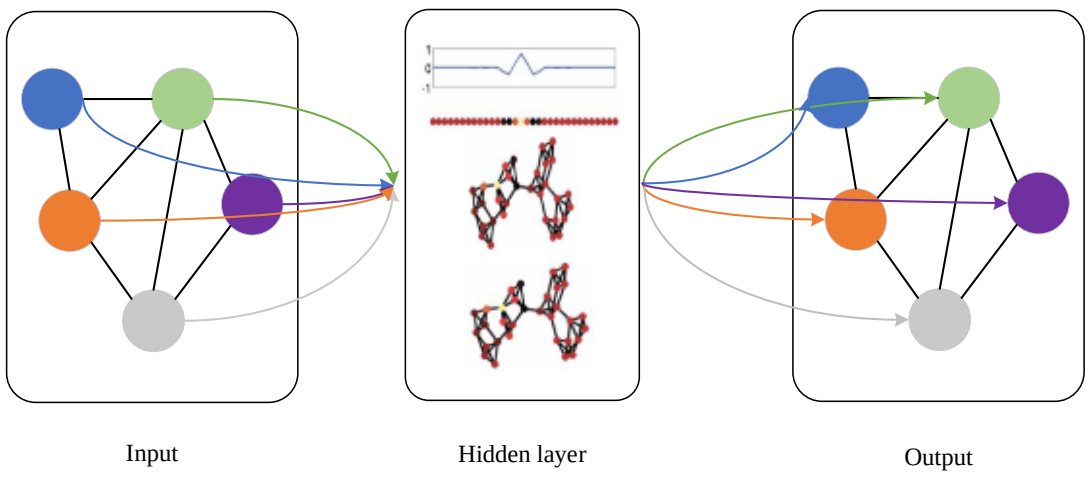

Fig. 3. Convolution of the spectrum wavelet in the frequency domain.

### 3.5. The DDIA spatio-temporal feature extraction

To simultaneously incorporate the DDIA temporal and topological features, we combine the spatio-temporal attention mechanism, the graph wavelet convolution layer, and the gated dilated causal convolution layer to construct a spatio-temporal graph convolution layer model. In general, a spatio-temporal graph convolution module consists of a stack of spatio-temporal attention layers, graph wavelet convolution layers, and gated dilated causal convolution layers. The specific structure's order and the number of connections are determined based on experimental results for the particular problem at hand. In this study, we primarily employ a sandwich structure, which includes a sequence of spatio-temporal attention layers followed by spatio-temporal graph convolution layers. This design incorporates a spatial layer (graph wavelet convolution layer) between two temporal layers (gated dilated causal convolution layer) to fuse the spatial graphs of power systems from multiple time slices. This allows us to mine the correlation of features in both the temporal and spatial dimensions of DDIA attacks.

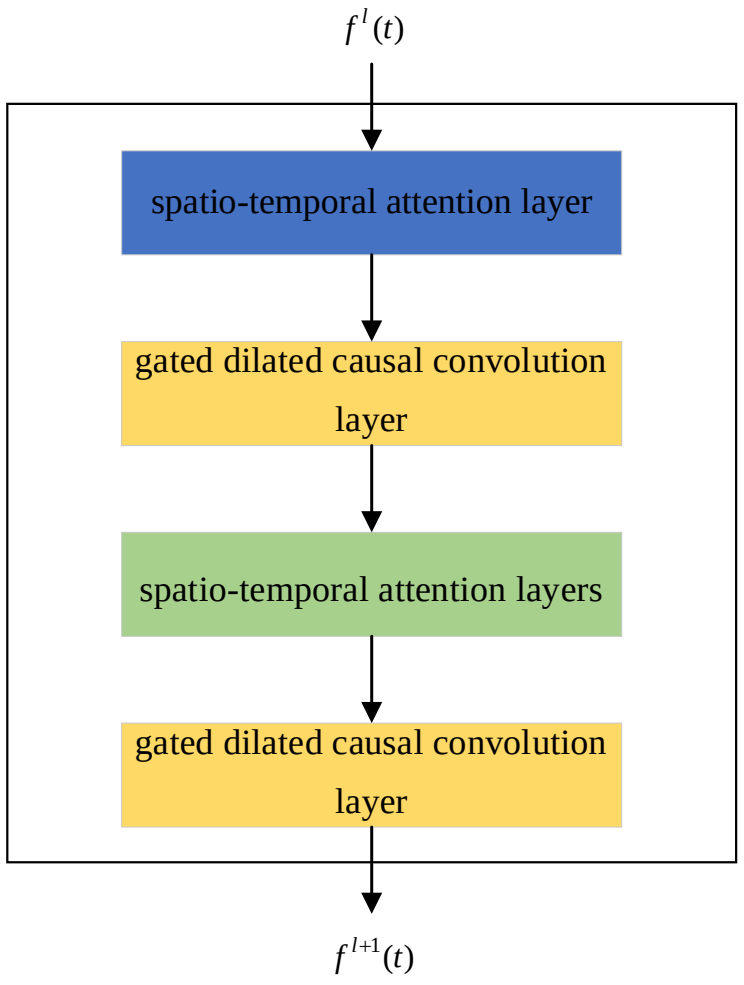

Fig. 4. Spatio-temporal graph convolution module.

It is worth noting that our proposed graph wavelet convolutional neural network differs significantly from the reference [40].

1) Combination of Graph Convolutional Neural Network and Wavelet Transform: Our approach integrates graph neural networks with wavelet transform in a unique manner, particularly suited for DDIA detection in power systems. Our network combines wavelet transform with convolutional computation to form an integrated feature extraction mode for feature extraction. Instead of performing wavelet transform first and then applying neural network classification. The graph wavelet transform is used to process the power system data, extracting both the frequency and time characteristics essential for identifying subtle attack patterns.

2) Methodological Distinctions: While it is true that neural networks combined with wavelet transforms have been previously employed in attack detection, our methodology differs from the reference [40] in several ways:

- Target of Detection: Our study focuses specifically on DDIAs in power systems, which presents unique challenges compared to the general false data injection attacks discussed in Yu et al.
- Spatio-Temporal Aspects: We incorporate a spatio-temporal aspect to our neural network model, which is particularly important in understanding the dynamics of power systems and the propagation of attacks within them.
- Graph Wavelet Approach: Our use of graph wavelet transforms, as opposed to traditional wavelet transforms, allows us to capture the topological structure of the power grid, which is crucial for detecting attacks that exploit the specific vulnerabilities of power system networks.

## 4. The DDIA localization method based on spatio-temporal graph neural network

From the attacker's perspective, the goal of the DDIA is to disrupt the stable operation of the power system. However, from the regulator's viewpoint, it is essential not only to promptly detect the occurrence of attacks but also to accurately identify the attacked bus's location. To achieve DDIA localization, we develop a comprehensive spatio-temporal graph neural network model that maps the power system measurement data to the attacked bus locations. Inspired by the concepts presented in references [37] and [38], we transform the DDIA localization problem into a multi-label binary classification task. In this approach, a binary label is assigned to each load bus, with a label value of 1 indicating the presence of an attack on that specific bus. Furthermore, an additional binary label is introduced to signify the presence of an attack on the entire power system network.

### 4.1. Input and output of the DDIA localization model

The operation of the power system exhibits differences between normal and post-attack states, with the measured data showing correlations in spatial topology and dynamic temporal patterns.

Temporally, the power system measurement data encompass dynamic time-series characteristics, including voltage magnitude, node active power injection, node reactive power injection, line active power, and line reactive power. To prevent overfitting during the localization model training, only vertex features within the power system graph model are considered as temporal inputs, specifically nodal active power and nodal reactive power.

In terms of spatial features, the inherent topological information of the power system can be represented using the network's admittance matrix. Consequently, the admittance matrix is employed as the spatial feature input for the model. It's important to note that the power system's topology can vary across different scales, and the non-Euclidean spatial nature of the admittance matrix can effectively capture the distinct spatial features of various power systems. This prior knowledge of the power system's inherent topological information bolsters the interpretability and credibility of the attack localization model.

The model's output indicates whether each node is under attack or not, providing information on the attack's location and the presence of an attack across the entire power system network.

### 4.2. Model structure design

The proposed attack localization model, referred to as the Attention Spatio-Temporal Graph Wavelet Convolutional Neural Network (ATSTGWCN), features an overall structure depicted in Figure 5. The model comprises an input layer, a hidden layer, a fully connected layer, and an output layer. Notably, the hidden layer represents a fusion of the DDIA spatio-temporal feature extraction module, serving as the central element for automatically extracting potential spatio-temporal features of the DDIA. To ensure that the training process remains stable and that attack localization accuracy is improved, a residual connection structure is incorporated, preventing a decrease in performance with deeper layers. The model is also equipped with normalization layers, a dropout mechanism, and other components to mitigate overfitting issues and enhance computational efficiency. The activation function chosen for the output layer is Sigmoid. Furthermore, all unknown parameters within the model are trained and learned using the cross-entropy loss function:

$$L(\hat{y}, W_\theta) = -\frac{1}{N}\sum_{n=1}^{N} y_i \log(\hat{y}_i) + (1-y_i)\log(1-\hat{y}_i) \quad (38)$$

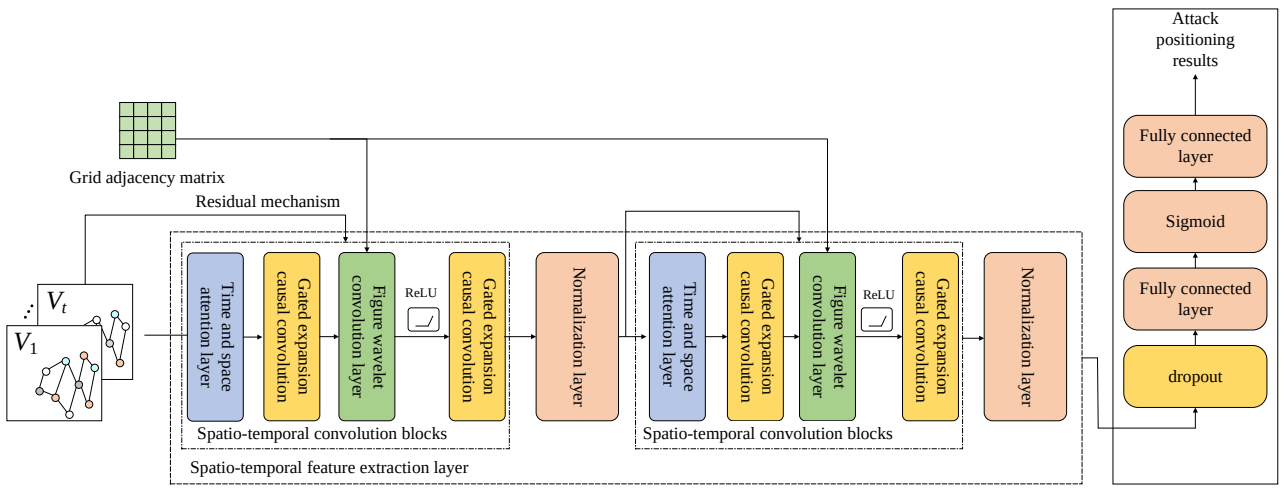

Fig. 5. Structure of ATSTGWCN model

### 4.3. The DDIA localization process

The process of DDIA localization using a spatio-temporal graph neural network is visualized in Figure 6. This flowchart illustrates the sequence of steps involved in the localization of DDIA through the utilization of a spatio-temporal graph neural network.

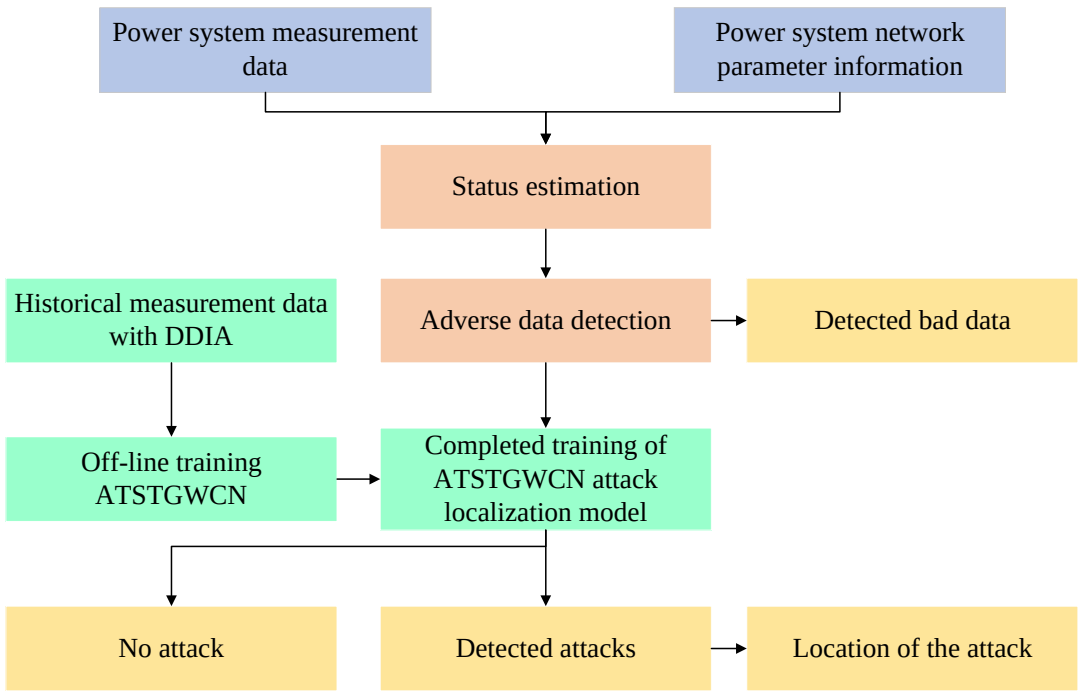

Fig. 6. The DDIA localization flow chart

The DDIA localization process involves three key stages:

1. Sample Generation with the DDIA: This phase begins with the construction of eligible DDIA models as outlined in Section 1 of the paper. These DDIAs are used to create complete datasets, which encompass both pre-attack and post-attack measurements. The temporal and spatial features extracted from these datasets are selected as inputs for the localization model.

2. Offline Training: In this stage, the entire dataset is divided into training and test sets. The long vector data generated from training set samples are transformed into a graph data format to facilitate processing by the localization model. The model parameters are then obtained through automatic training and learning using the training data to find the best-fitting model.

3. Online Localization: The test dataset is provided as input to the trained model for online localization. The model processes this input and generates localization results for the DDIA based on the learned parameters.

These three stages together form the comprehensive process of DDIA localization, enabling the model to detect and pinpoint the presence of DDIA attacks within a power system.

## 5. Experimental results

This paper evaluates the proposed method on three distinct power systems of varying sizes: IEEE 14-node, IEEE 118-node, and IEEE 300-node systems. The network topology, measurement data, and parameter information for these systems are obtained from the Matpower toolbox. For conducting simulation experiments, the machine configuration used includes an Intel Core i7-7700 CPU, two NVIDIA GTX 1080 GPUs, and 64GB of RAM. This setup ensures the effective and rigorous testing of the proposed method's performance and superiority in detecting and localizing DDIAs in power systems.

### 5.1. Dataset Generation

In the absence of publicly available datasets due to data privacy constraints in the power system, the measurement data for normal system operation and the dataset after DDIAs were generated using the DDIA model developed in the previous section. The datasets were generated for three different power systems: IEEE 14-node, IEEE 118-node, and IEEE 300-node systems. Each of these systems has a distinct configuration, including the number of generators and branch circuits.

The collected measurement data consists of node voltage magnitude, node active and reactive power, and branch active and reactive power. The process of dataset generation involves the following steps:

1. Normal Measurement Data: Generate system load data following a normal distribution with a mean value equal to the base load and a standard deviation of 1/6 of the base load. Complete measurement data is obtained through a power flow operation.

2. Measurement Data After Being Attacked: Optimize the injection of dummy data using a linear programming approach. Each working condition corresponds to an optimal solution. To create an attack sample, simulate a random attack optimizing each working condition. Based on the DDIA model constructed earlier, generate the attacked measurement data. It's important to note that generator output values and all measurement data in the non-attack area remain unaltered.

3. Training and Test Set Division: Create a total of 20,000 normal measurement samples and 20,000 attacked measurement samples. Out of these, 15,000 normal samples and 15,000 attacked samples are uniformly selected for the training set. The remaining samples are reserved for the test set. To ensure robustness, a 5-fold cross-validation approach is employed.

This data generation procedure ensures that you have a comprehensive dataset for evaluating the performance of the proposed DDIA detection and localization method across different power system configurations and attack intensities.

### 5.2. Validation of the DDIA model

The paper includes an assessment of the DDIA model's accuracy and impact on the power system, specifically evaluating state changes before and after attacks. This analysis involves an examination of the power system's overload conditions and how the success rate of the attack is influenced by the degree of topology integrity.

To verify the effectiveness of the DDIA model, the study simulates an attacker with partial topology information. The attacker can manipulate measurement values in specific regions, in accordance with the consistency constraints of Kirchhoff's Voltage Law (KVL) and Kirchhoff's Current Law (KCL). In this simulation, a medium attack strength and an 80% mastery of topology are assumed. The Monte Carlo method generates 50 sets of normal measurements.

In Figure 7, a comparison is made between the numerical changes in the IEEE 14-node system under normal conditions, estimated state, and post-attack state. After the DDIA is initiated, the voltage magnitude and voltage phase angle, which characterize the power system's operational state, deviate to different degrees from their expected values as per state estimation. The post-attack state illustrated in the figure demonstrates the potential for misleading regulators and jeopardizing the safe operation of the power system. This analysis underscores the significant impact of DDIA on power system operation.

1) Clarification on Constraint (10): The constraint specified in Equation (10) intends to ensure that the voltage amplitudes at the boundary nodes (region B) remain consistent before and after the attack. This constraint is crucial to make the attack stealthy by not causing noticeable changes in the boundary nodes, which are more likely to be closely monitored in a power system.

2) Explanation of Figure 7 Results: The deviations observed in Figure 7 primarily reflect the voltage amplitudes at the internal nodes within the attack region, rather than at the boundary nodes. These internal nodes are where the attackers have more freedom to manipulate data without being easily detected. The large deviations you noted are a result of the attackers' actions within the region they control, which do not necessarily violate the constraint of maintaining voltage consistency at the boundary nodes.

3) Reconciling the Results with the Constraints: We understand that the observed large deviations in the internal nodes of the attack region might give an impression of contradicting the constraints. However, these results align with our model's intention to demonstrate the potential impact of an attack within the confines of the constraints. The attack is designed to be subtle at the monitored boundaries (region B) while being more aggressive internally, which is what Figure 7 illustrates.

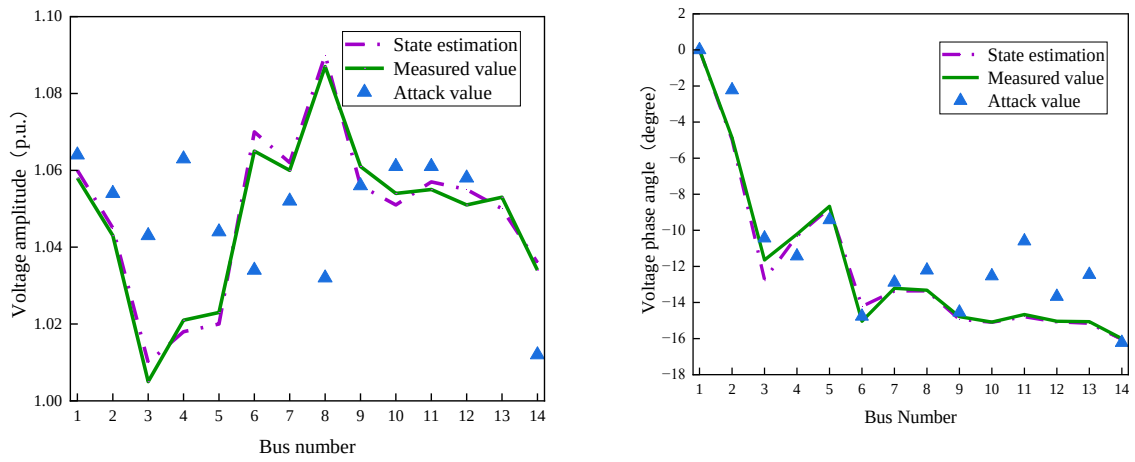

Fig. 7. State change before and after the attack

To demonstrate the covert and destructive nature of the constructed DDIA, the three test systems are visualized using PCA (Principal Component Analysis) dimensionality reduction. This technique provides a two-dimensional representation of the normal measurement data and the measurement data after an attack, as depicted in Fig. 8. Additionally, the paper provides tables (Tables 1-3) showing power system line overloads resulting from the DDIA attacks. These visualizations and data tables help illustrate the impact of the DDIA on the tested power systems, highlighting its ability to disrupt and remain hidden within the normal operation data.

Regarding the similarity of results depicted in our Figure 8 and Figure 2 of Liu et al. [8], we acknowledge that certain attack scenarios might produce similar outcomes, particularly when the attacks are designed with similar objectives or methodologies. However, the context and interpretation of these results are unique to our study, given our focus on the integration of spatio-temporal aspects and graph wavelet convolutional methods, which are not addressed in Liu et al.'s study. In addition, our proposed DDIA model is based on incomplete topological information and AC power flow equations, which is different from current research.

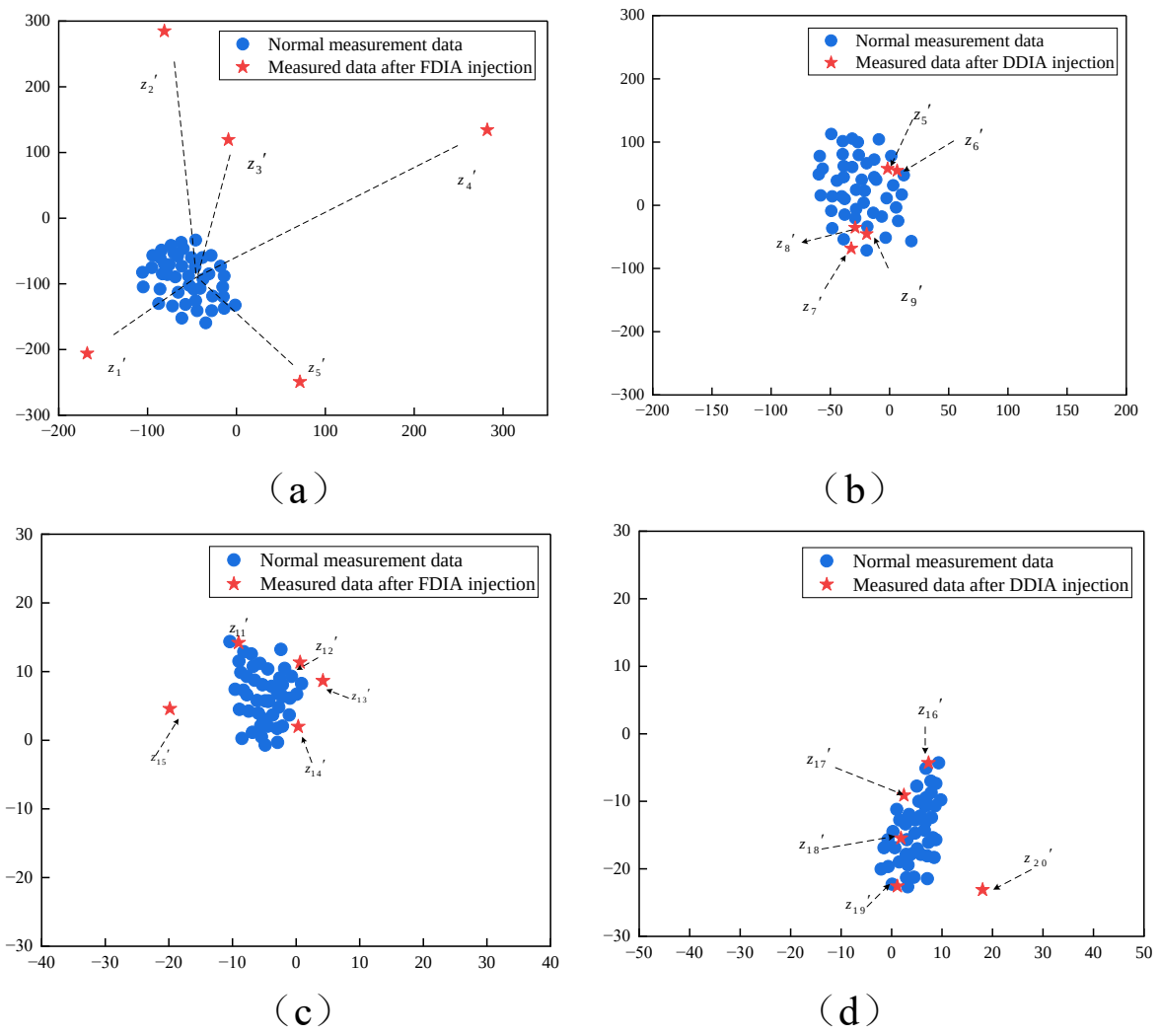

Fig. 8. Effect of the DDIA injection test system

Tab. 1.
Line overloads in IEEE 14-node systems caused by the DDIA

| Overload line | $F^l / F^l_{max}$ Overload Rating | Distance Index |
|---|---|---|
| Line 2-3 | 1.43 | 40.45 |
| Line 5-6 | 1.42 | 36.82 |
| Line 6-13 | 1.42 | 0.68 |
| Line 9-14 | 1.41 | 5.66 |
| Line 12-13 | 1.44 | 10.89 |

Tab. 2.
Line overloads in IEEE 118-node systems caused by the DDIA

| Overload line | $F^l / F^l_{max}$ Overload Rating | Distance Index |
|---|---|---|
| Line 7-12 | 1.44 | 30.24 |
| Line 17-18 | 1.43 | 28.53 |
| Line 25-27 | 1.42 | 35.66 |
| Line 49-54 | 1.45 | 25.41 |
| Line 92-102 | 1.42 | 86.47 |

Tab. 3
LINE OVERLOAD OF IEEE 300-NODE SYSTEM CAUSED BY THE DDIA

| Overload line | $F^l / F^l_{max}$ Overload Rating | Distance Index |
|---|---|---|
| Line 5-9 | 1.46 | 37.67 |
| Line 11-13 | 1.42 | 25.87 |
| Line 81-88 | 1.41 | 13.88 |
| Line 116-124 | 1.43 | 12.64 |
| Line 195-212 | 1.44 | 78.68 |

Fig. 8(a) demonstrates the impact of a traditional FDIA injected into the IEEE 14-node system, while Fig. 8(b) illustrates the effects of a DDIA injected into the same system. Fig. 8(a) shows that traditional FDIA does not consider the spatial distance's similarity features. As a result, the distances between the constructed false data and normal data appear random and widely dispersed. In this figure, you can see that twenty measurements are gathered closely, all indicating safe states, whereas the five isolated false data points represent unsafe states. The distance index between measurements is calculated based on a spatial rectangular coordinate system, with the distance index range computed as [0, 28.23] based on the set of normal measurements. However, in the case of traditional FDIA, the distances after the attack ( $z_1'$ , $z_2'$ , $z_3'$ , $z_4'$ , and $z_5'$ ) are 58.45, 64.72, 71.88, 75.86, and 93.87, respectively. These values are significantly larger than the distance index of the normal measurements threshold, making this attack conspicuous and easily detectable using distance-based detection methods. Table 1 reveals that these five sets of dummy data relate to measurements on lines 1-2, 6-10, 8-9, and 8-10, all leading to an overload situation with an overload level 1.4 times the normal line trend threshold.

In contrast, Fig. 8(b) shows that the distances between the DDIA-attacked measurements and the normal measurements are very close. Despite their proximity, the DDIA causes overloads in the power system. The distance index range for normal data in Fig. 8(b) is [0, 50.35]. In the case of the DDIA, the distances after the attack measurements ( $z_5'$ , $z_6'$ , $z_7'$ , $z_8'$ , and $z_9'$ ) are 40.45, 36.82, 0.68, 5.66, and 10.89, respectively. These results highlight that the DDIA-attacked data are concealed within the normal measurements, satisfying the DDIA's concealment feature and rendering the distance detection method invalid.

Fig. 8(c) and Fig. 8(d) depict the effects of the DDIA injected into the IEEE 118-node system and the IEEE 300-node system, respectively. Similar to the IEEE 14-node system, the DDIA in both cases is hidden within normal measurement data, evading many Euclidean distance-based detection methods. These DDIA attacks lead to overloads in the respective lines, indicating their destructive nature. It's worth noting that not all attacked data can blend with normal measurement data and go undetected. As seen in the figures, the distance indices of $z_{15}'$ and $z_{20}'$ are higher than the limited range, meaning that the success rate of the DDIA does not reach 100%. Nevertheless, this doesn't negate the fact that all these data have a destructive impact on causing line overloads. It simply indicates that the DDIA becomes more similar to the original FDIA.

By examining Fig. 8(a), (b), (c), (d) and Tables 1, 2, and 3, it is evident that the simulation experiments have successfully constructed the DDIA, which is both well-concealed and harmful to the power system.

To investigate the impact of the attacker's degree of topology mastery on the attack's success rate, the success rate is defined as the ratio of the number of successful bypasses of existing bad data detections to the total number of attacked data, as shown in Fig. 9.

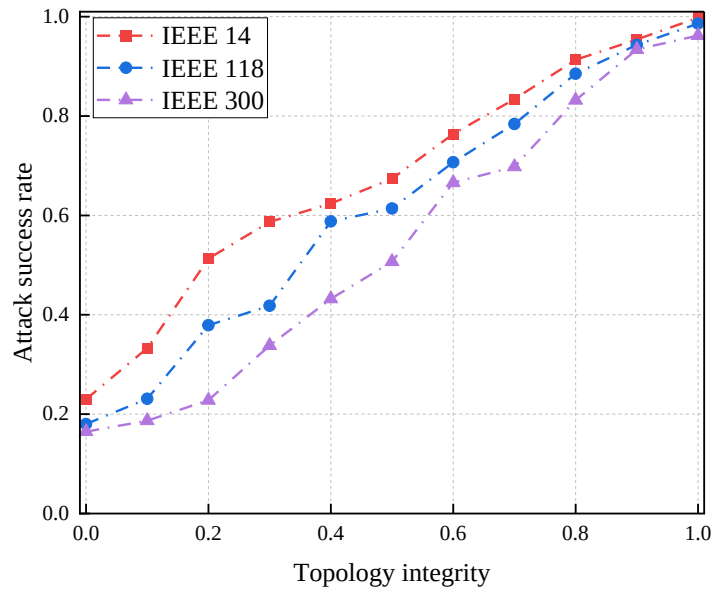

Fig. 9. Effect of topology integrity on the success rate of the attack

Fig. 9 reveals a correlation between the attack's success rate and the attacker's level of topology mastery. As the attacker's knowledge of the power system's topology increases, the success rate of the attack tends to rise. When the attacker is entirely uninformed about the power system's topology, it falls under the category of a blind information attack, resulting in a low success rate. Conversely, when the attacker has full knowledge of the topology, the success rate approaches 1. The level of uncertainty in the topology and the system's scale do have varying effects. The attack's success rate varies depending on the specific topology and size of the system. For instance, when the topology integrity is at 0.7, the success rate for attacking the IEEE 14-node system is 0.832, for the IEEE 118-node system it's 0.788, and for the IEEE 300-node system, it's 0.746. This illustrates how the uncertainty and larger scale of the power system's topology can influence the success rate of attacks. It also underscores that larger-scale systems are more resilient to interference and demonstrate higher resistance to attacks. From a defender's perspective, it is crucial to maintain the confidentiality of topology information and line parameters to prevent data leakage.

The comparative experiments discussed above underscore the effectiveness and accuracy of the DDIA model. These attacks can lead to severe consequences in the power system, including line overloads. The attacker's level of topology information mastery in DDIA significantly impacts the attack's success rate.

### 5.3. The DDIA localization model performance

The DDIA localization model's training and learning process employs an Adam optimizer with a learning rate set to 0.0001. During training, a random dropout strategy is implemented with a dropout rate of 0.2 and a scale factor of s=2. The cross-entropy loss function is computed using the binary cross-loss entropy described in equation (36). Training samples are fed to the localization model in small batches of 64, and the training process consists of 100 rounds. The input data for the nodes includes active power, reactive power, and the admittance matrix.

The performance of the localization models is evaluated using various metrics such as the number of true positive samples, true negative samples, false positive samples, and false negative samples, which were calculated through simulation experiments. Performance evaluation metrics include accuracy (AC), precision (PR), recall (RE), false alarm rate (FA), and F1 score, as defined in the related reference [39]. During the classifier's prediction process, n+1 label values are returned at each time step, and the mean value of the prediction performance is chosen as a metric for comparing the performance of each localization model.

The tests were conducted on a three-node system with a moderate attack intensity. This paper compares various machine learning and deep learning models, including Support Vector Machine (SVM), K-Nearest Neighbor (KNN), Decision Tree (DT), Random Forest (RF), Light Gradient Boosting Machine (LightGBM), CNN, RNN, Long Short-Term Memory Network (LSTM), Graph Convolutional Neural Network (GCN), and Spatio-Temporal Graph Convolutional Neural Network (STGCN). Additionally, comparisons are made with existing methods such as Extreme Learning Machine (ELM) [20] and Improved Capsule Network (ICN) [24], as well as Graph Neural Network with Attention Mechanism (ATSTGAT).

(1) Training loss and testing accuracy of the localization model

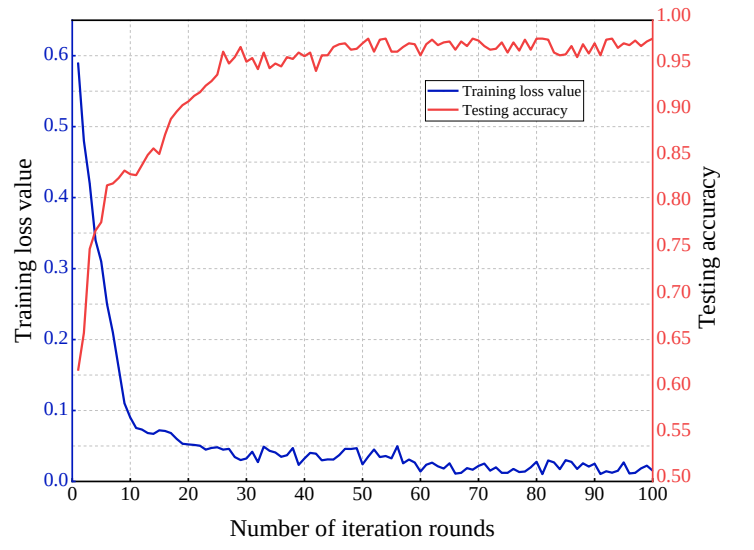

Fig. 10. Loss value and accuracy variation during training and testing

Fig. 10 illustrates the changes in loss value and accuracy with the number of iteration rounds during the training and testing process of the IEEE 14-node system. As the number of iteration rounds increases, the overall trend of the training loss value decreases, and convergence is achieved more rapidly. The loss function approaches convergence after approximately 30 rounds, stabilizing between 0.02 and 0.03 when training reaches around 60 rounds.

This suggests that the model exhibits efficient computation and can rapidly extract the spatio-temporal features of the DDIA.

The test accuracy of the model also improves with each iteration. After 60 rounds, the accuracy reaches a level between 0.96 and 0.98, indicating that the model can accurately detect and locate the attack's position. The training speed and testing accuracy demonstrate that the localization model can quickly and effectively identify attacks, which is of great significance for defenders to take timely and appropriate measures.

(2) Performance comparison of the localization model and other models

Fig. 11 presents the results of the ATSTGWCN model proposed in this paper compared with other benchmark localization models using the IEEE 14-node system. The results indicate that the ATSTGWCN attack localization model has the highest values for AC, PR, RE, and F1 metrics, reaching 0.9758, 0.9789, 0.9732, and 0.97574, respectively. This demonstrates that the model effectively enhances attack localization by considering spatio-temporal correlations in the attack data and focusing on local neighborhood features. The FA metric also reflects the lowest value of 0.0216, indicating the superior performance of the proposed model.

In the comparison, different models are categorized into distance-based, tree-structured, neural network-based, and graph convolution-based models. The model proposed in this paper is based on STGCN, which captures topological correlation features but suffers from issues related to Chebyshev polynomial approximation. ATSTGAT, another high-performing model, considers both temporal and spatial aspects but may face problems such as gradient explosions and unsatisfactory local feature extraction. The proposed model combines spatio-temporal attention mechanisms, gated time convolution, and graph wavelet spatial convolution, providing improved attack feature extraction adaptability, long sequence correlation extraction capabilities in the time domain, and spatial feature extraction capabilities.

Similar performance comparison experiments were conducted on the larger-scale IEEE 118-node and IEEE 300-node systems. Although deep learning models may not necessarily outperform machine learning models like RF and LightGBM due to the risk of overfitting, ATSTGWCN, which combines gated time convolution and graph wavelet convolution, overcomes such problems and performs well. This demonstrates the adaptability of the proposed localization model to power systems with varying topologies and its capacity to deliver satisfactory localization results, even in large-scale systems.

While this paper primarily compares single models such as SVM, KNN, DT, RF, LightGBM, CNN, RNN, LSTM, and GCN, it also highlights the success of the GCN model. Experimentally, the study indicates that layering additional modules on top of the GCN model is effective. It also confirms the advantage of the proposed model in localization performance compared to the ATSTGAT model.

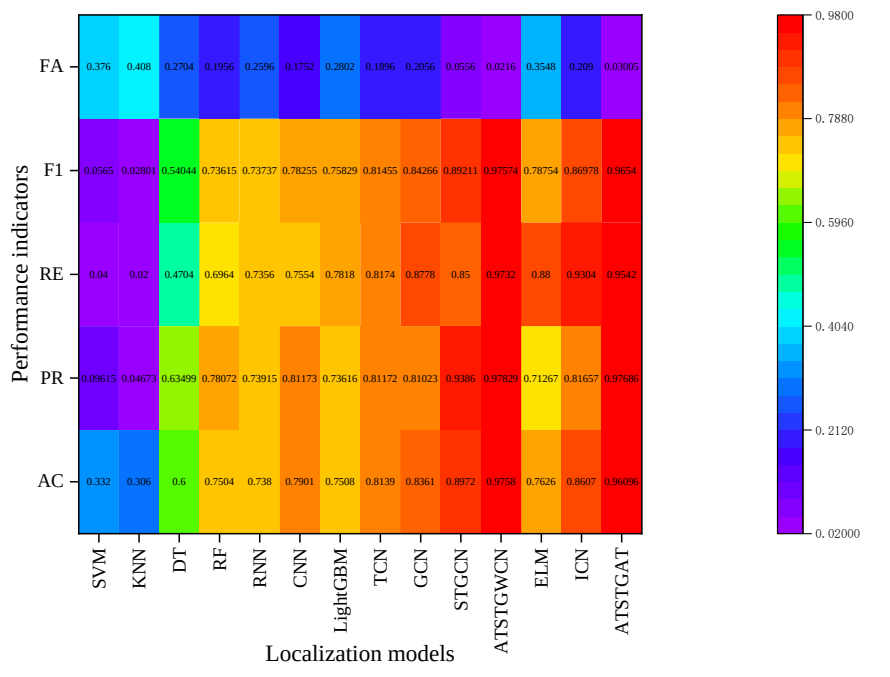

（a）IEEE 14

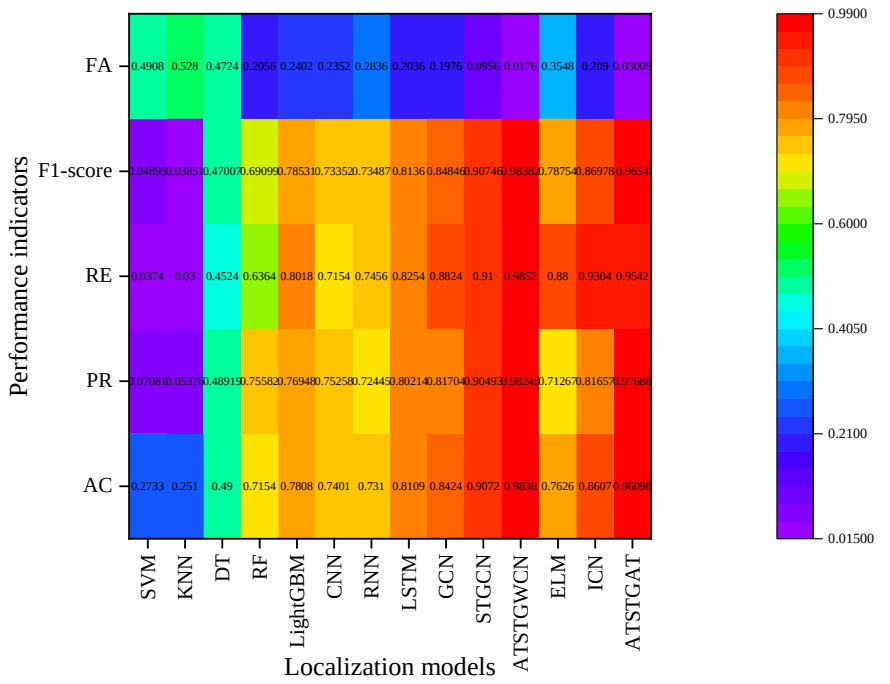

（b）IEEE 118

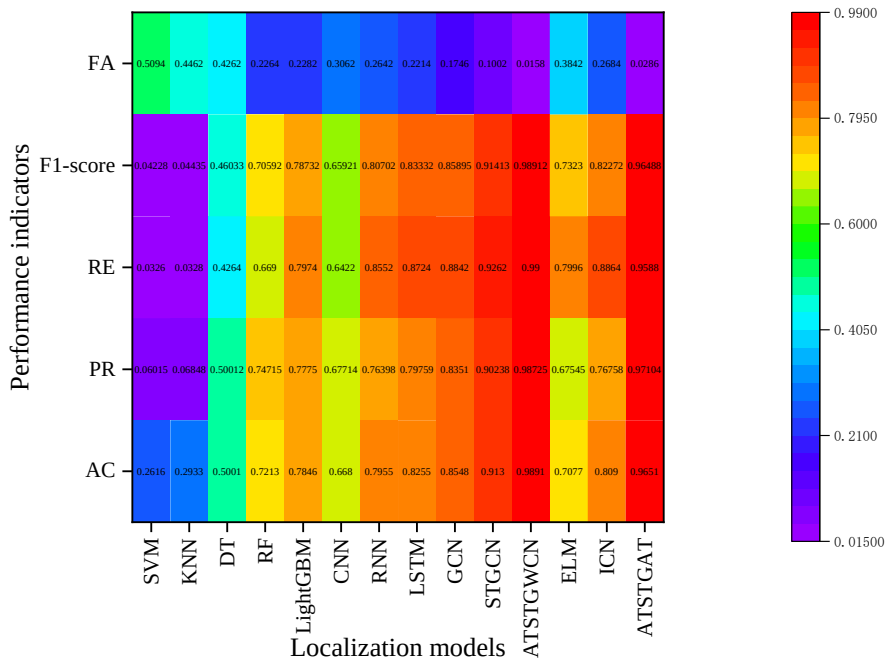

（c）IEEE 300

Fig. 11. Comparison of the results of ATSTGWCN model with other benchmark localization models

(3) Receiver operating characteristic (ROC) curve of the model

Fig. 12 presents ROC curves of different localization models for the IEEE 14-node system using 3D waterfall plots, which help visualize performance under different hyperparameters. The degree of convexity of the curve toward (0,1) reflects the superiority of localization performance. Generally, a larger area under the curve (AUC) indicates better model performance. It's evident that ATSGWCN has the largest AUC area, signifying the best performance in attack localization, followed by STGCN and GCN.

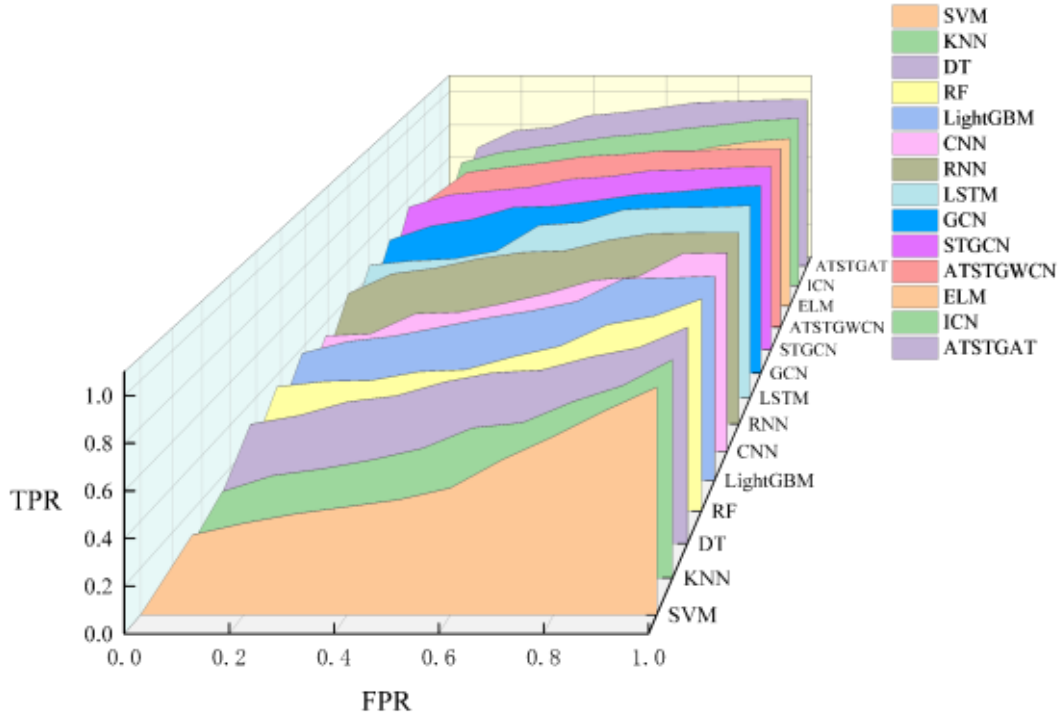

Fig. 12. ROC curves for different localization models of IEEE 14-node system

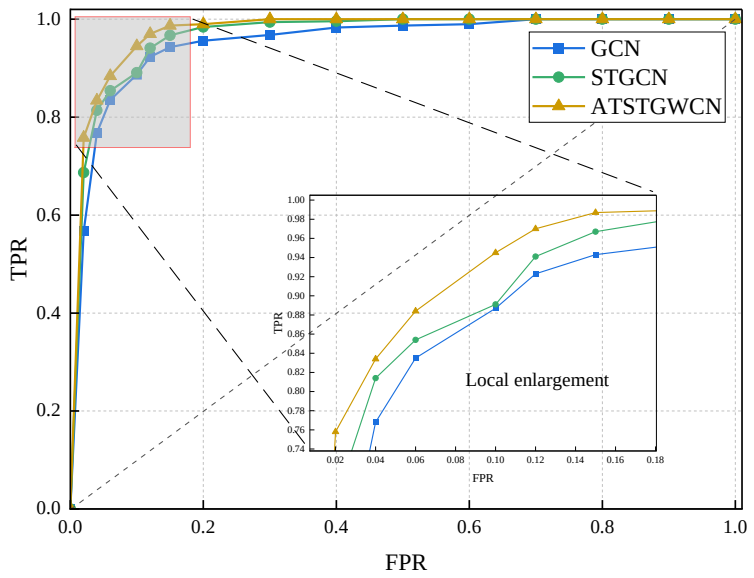

Fig. 13. ROC curves for GCN, STGCN, and ATSGWCN for the IEEE 118-node system

Fig. 13 focuses on the comparison of ROC curves for GCN, STGCN, and ATSGWCN in the IEEE 118-node system. The results show that the model proposed in this paper still outperforms others in large-scale systems. These ROC curves further emphasize the superior localization performance of the ATSGWCN model.

(4) Comparison of localization performance under different attack strengths

The preceding section demonstrated the advantages of the proposed attack model. This section quantitatively analyzes the impact of attack strength on the model's localization performance. Attack intensities are categorized as strong, medium, and weak, and experiments are conducted across all three systems. The comprehensive F1-score is chosen as the evaluation metric, as depicted in Fig. 14.

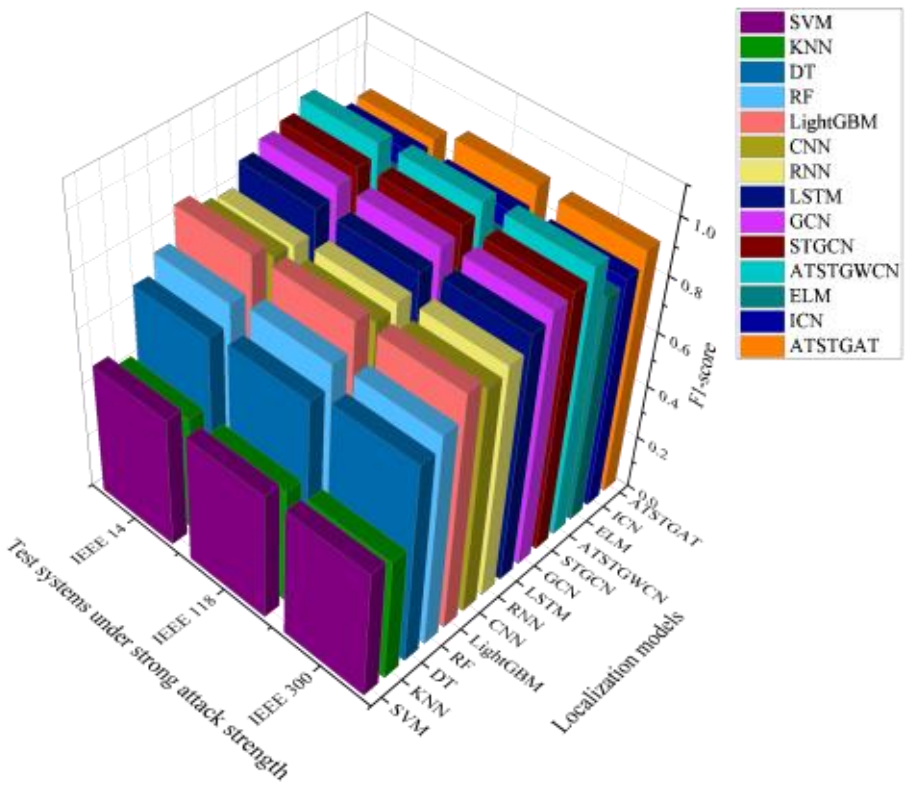

(a) Model localization performance under strong attack strength

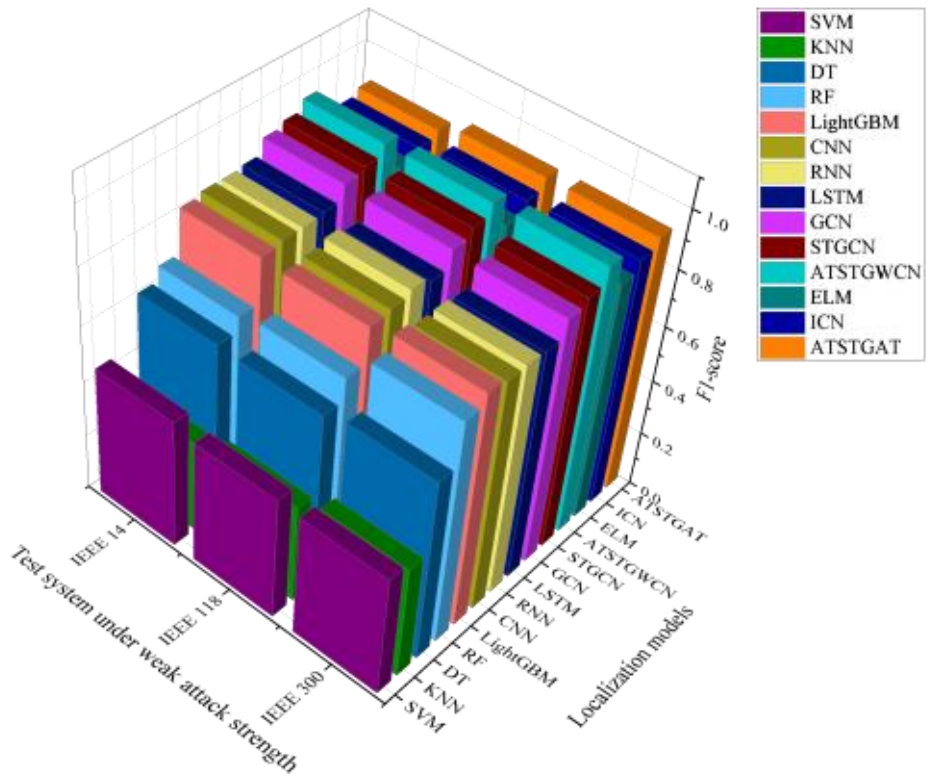

(b) Model localization performance under weak attack strength

Fig. 14. Impact of attack intensity on model localization performance

In the case of strong attack intensity, most localization models, except distance-based SVM and RNN models, show improved performance. This is because the attack induces significant fluctuations in voltage magnitude, node power, line flow, and other measurement data, leading to increased differences in class distribution between samples and making attack features easier to capture. Under strong attack strength, the F1-scores for the proposed localization model are 0.9767, 0.9815, and 0.9843 when tested in IEEE 14, IEEE 118, and IEEE 300 systems, respectively, indicating clear advantages and outperformance compared to other models.

ATSTGAT, STGCN, ICN, and GCN follow as the second-best performers, while LightGBM, CNN, RNN, LSTM, and ELM exhibit similar values. The remaining models have lower F1-scores, making it challenging to accurately locate DDIAs. As the number of nodes increases, the spatial correlation of local neighborhood features in measured data becomes stronger. Spectral transformation with wavelet convolution reduces operational complexity and enhances feature extraction capability. Moreover, the incorporation of gated time convolution allows for more robust capture of the temporal correlation in long measurement sequences. For weak attack intensity, the F1-scores of the three systems in the localization model are 0.9573, 0.9588, and 0.9612, respectively. Although there is a performance decrease for weak attack intensity, it remains acceptable.

(5) Effect of the number of training set samples on the localization performance

In machine learning, the number of training samples often has a significant impact on the performance of a classifier's predictions. To verify this impact, the total number of training samples is varied between 2000, 4000, 6000, 8000, 10000, and 12000, with an equal balance of attack and normal samples (1:1 ratio) in each training set. The number of test samples remains the same, and the F1 scores of the ATSTGWCN model proposed in this paper are shown in Fig. 15.

The overall trend of F1 scores for the three systems increases as the number of training samples increases. F1 scores show significant improvement between 2000-8000 samples and gradually stabilize between 8000-14000 samples. This suggests that the number of training samples indeed affects model performance, and increasing the number of training samples can moderately enhance localization performance. However, a larger training set does not necessarily translate to better test performance. When the number of training samples reaches 10,000, the F1 score stabilizes.

The corresponding confusion matrix is also provided in the figure, illustrating that the choice of training set size should be made judiciously based on the specific scenario. In other words, when it comes to attack localization, the training set size should not be too large, as it would waste computation time, nor too small, as it would adversely affect localization performance.

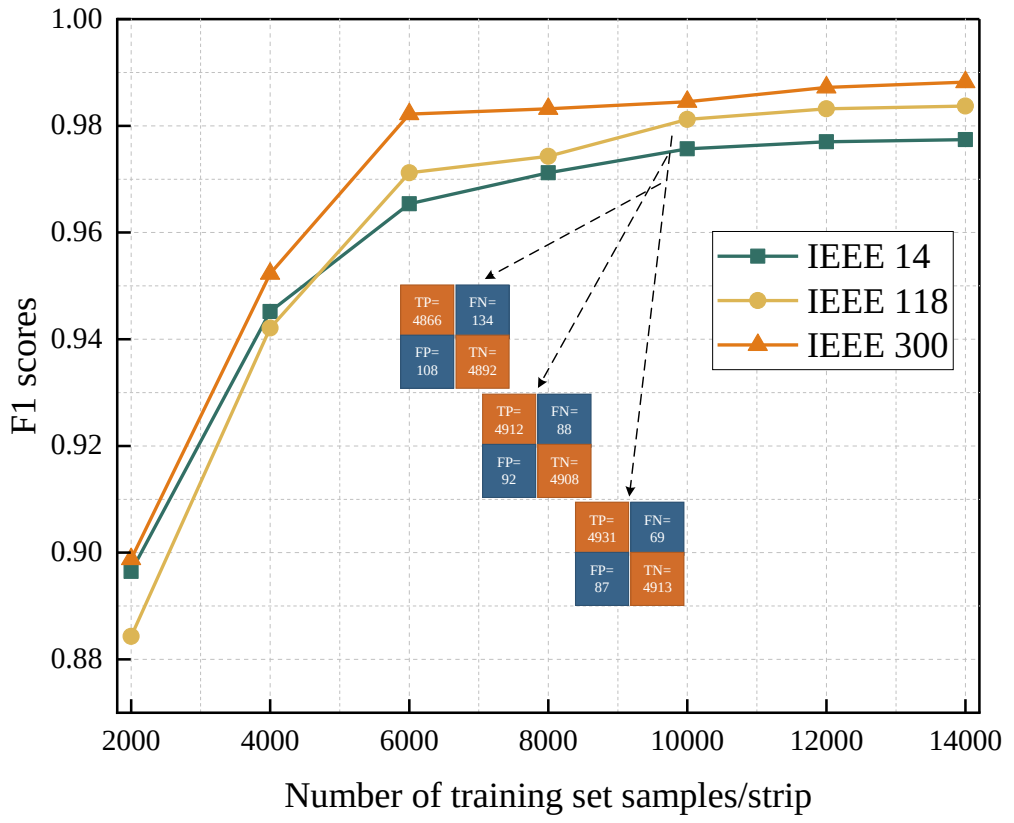

Fig. 15. Effect of the number of samples in training set on the localization performance

### 5.4. Advantages of the model structure

The study focused on the IEEE 14-node system, and Fig. 16 presents an experimental comparison of various metrics when the spatio-temporal attention mechanism is added and when it is not included. The results clearly indicate that the addition of the spatio-temporal attention mechanism improves the model's performance compared to the scenario without it. This improvement is attributed to the spatio-temporal attention mechanism's ability to adaptively extract crucial spatio-temporal information by assigning more importance to key features. This adaptability is beneficial for processing nonlinear and complex power system measurement data, ultimately reducing the error in attack prediction.

Furthermore, Fig. 17 illustrates the correlation strength between each node, as calculated by the spatio-temporal attention mechanism. In this visualization, darker colors in the rectangular blocks indicate stronger correlations between nodes. This graphical representation demonstrates that the proposed method in this paper not only enhances attack localization performance but also offers advantages in terms of interpretability and visualization.

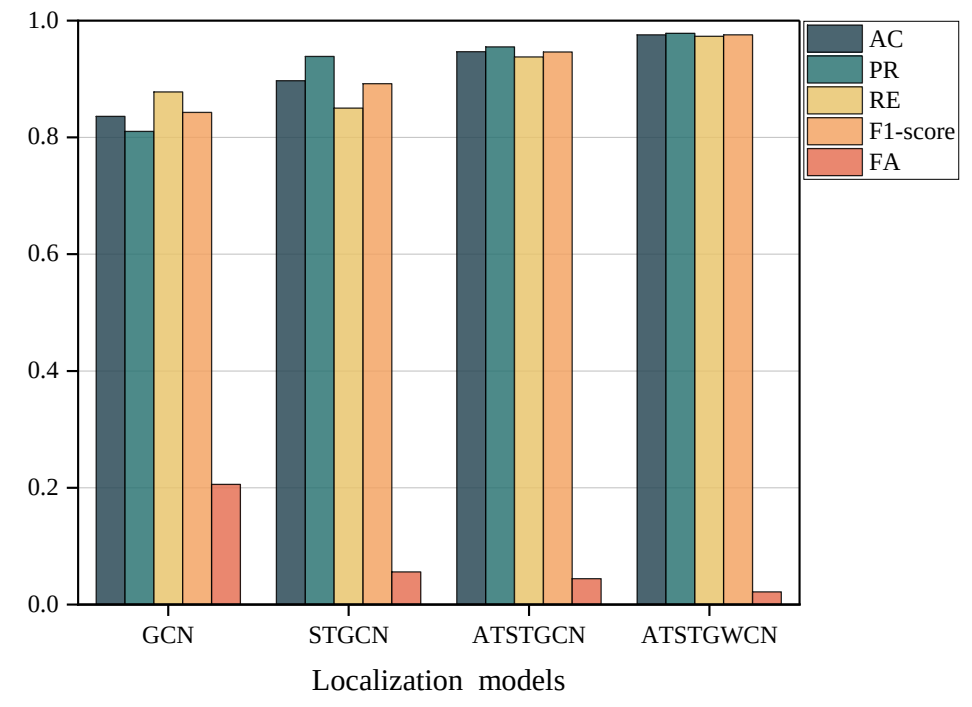

Fig. 16. Effect of attentional mechanism on localization performance

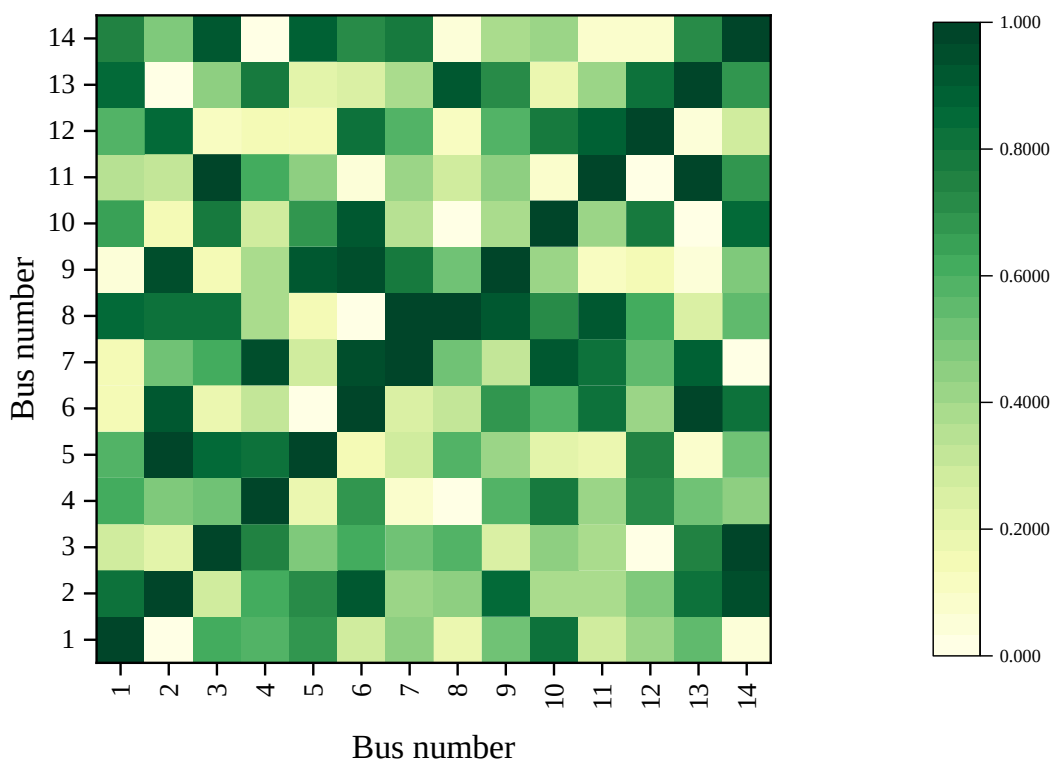

Fig. 17. Inter-node correlation strength

To assess the benefits of gated dilated causal convolution, various time prediction scales were considered to compare performance with the time-dependent benchmark model, as presented in Table 4. Since SCADA measurement data acquisition typically occurs at intervals of 2 to 5 seconds, the selected time scales were 10 minutes, 20 minutes, and 30 minutes to create sufficiently long time series. The localization performance varied across different time prediction scales, and the model proposed in this paper consistently outperformed the benchmark models across various metrics.

The inclusion of the dilated causal convolution and gating mechanism becomes more prominent and effective in processing time-series data as the time scale increases. This mechanism allows the proposed model to perform well even at the 20-minute time scale. However, for RNN, LSTM, and STGCN, as the time scale grows, their localization performance diminishes due to the increased computational burden associated with longer time series data.

The model proposed in this paper retains its strong ability to capture spatio-temporal features due to the dilated causal convolution and gating mechanism. The slight dip in performance at the 30-minute time scale, while not significant, suggests that the choice of time prediction scale should align with the specifics of the power system test environment to ensure optimal performance.

Tab. 4.
PERFORMANCE COMPARISON OF LOCALIZATION MODELS WITH DIFFERENT TIME PREDICTION SCALES

| | 10min | | | | | 20min | | | | | 30min | | | | |
|---|---|---|---|---|---|---|---|---|---|---|---|---|---|---|---|
| Models | AC | PR | RE | F1 | FA | AC | PR | RE | F1 | FA | AC | PR | RE | F1 | FA |
| RNN | 0.7380 | 0.7391 | 0.7356 | 0.7374 | 0.2596 | 0.7512 | 0.7716 | 0.7136 | 0.7415 | 0.2112 | 0.6891 | 0.7128 | 0.6334 | 0.6708 | 0.2552 |
| LSTM | 0.8361 | 0.8102 | 0.8778 | 0.8427 | 0.2056 | 0.7771 | 0.7554 | 0.8196 | 0.7862 | 0.2654 | 0.7633 | 0.7485 | 0.7930 | 0.7701 | 0.2664 |
| STGCN | 0.8972 | 0.9386 | 0.8500 | 0.8921 | 0.0556 | 0.8691 | 0.8939 | 0.8376 | 0.8648 | 0.0994 | 0.8412 | 0.8437 | 0.8376 | 0.8406 | 0.1552 |
| ATSTGWCN | 0.9758 | 0.9783 | 0.9732 | 0.9757 | 0.0216 | 0.9771 | 0.9762 | 0.978 | 0.9771 | 0.0238 | 0.9586 | 0.9597 | 0.9574 | 0.9585 | 0.0402 |

The comparison of computational complexity for spatial domain features, focusing on matrix sparsity, was conducted for the three test systems. The results are displayed in Table 5. It is evident that the graph wavelet frequency domain transform exhibits higher sparsity, leading to a reduced computational complexity for attack localization. This increase in sparsity not only speeds up the computation but also enhances feature extraction in the local neighborhood space, making it a more efficient and effective approach for attack localization.

Tab. 5.
COMPARISON OF THE SPARSITY OF GRAPH WAVELET TRANSFORM AND FOURIER TRANSFORM

| Percentage of non-zero elements | Wavelet transform $\psi^{-1}$ | Fourier transform $U^T$ |
|---|---|---|
| IEEE 14 | 25.87% | 98.85% |
| IEEE 118 | 20.96% | 98.04% |
| IEEE 300 | 19.77% | 98.21% |

In the graph wavelet convolution operation, the scale factor $s$ plays a crucial role in controlling the range of the diffusion neighborhood for graph signal features. Selecting an appropriate scale factor is essential for improving attack localization. The F1-score values for different node localization attacks were tested with various scale factors $s = 1, 2, 4, 8, 16$. As depicted in Fig. 18, the choice of scale factor indeed impacts the F1-score values. However, it's important to note that bigger $s$ does not necessarily translate to better performance. The selection of the neighborhood range size is a crucial consideration for improving model performance. The results in Fig. 18 indicate that, on average, a scale factor of $s = 2$ performs better than other choices.

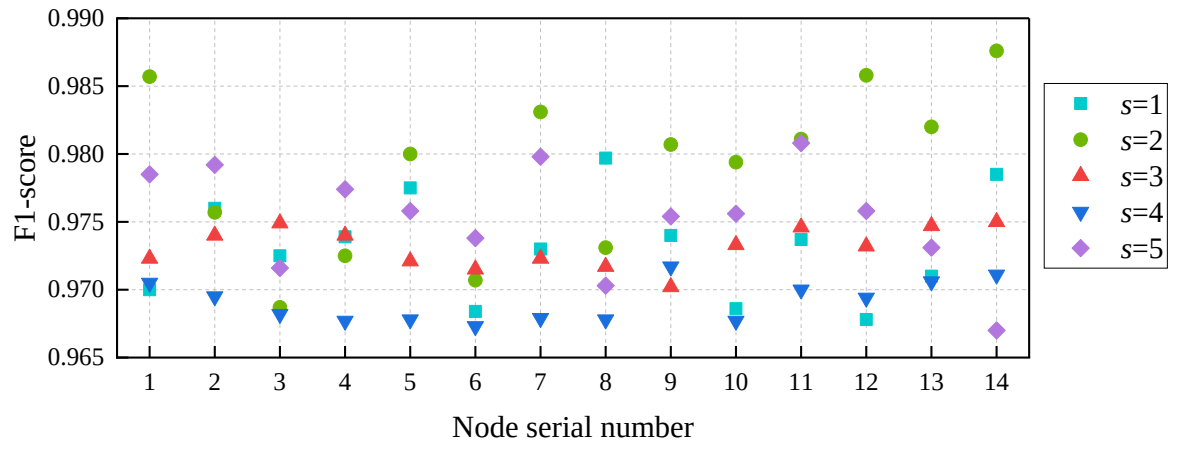

Fig. 18. Effect of scale factor on localization performance

The spatio-temporal graph convolution module designed in this paper follows a sandwich structure, involving a sequence of time-space-time convolutions. This design is beneficial for enhancing the model's attack localization performance. As depicted in Fig. 19, this structure comparison contrasts the time-space convolution module (T-S), space-time space convolution module (S-T-S), and the time-space-time convolution module (T-S-T) proposed in this paper. It also compares the attack localization capabilities of a single space-time graph convolution module (Sin) with three space-time graph convolution modules (Tri) with the model proposed in this paper.

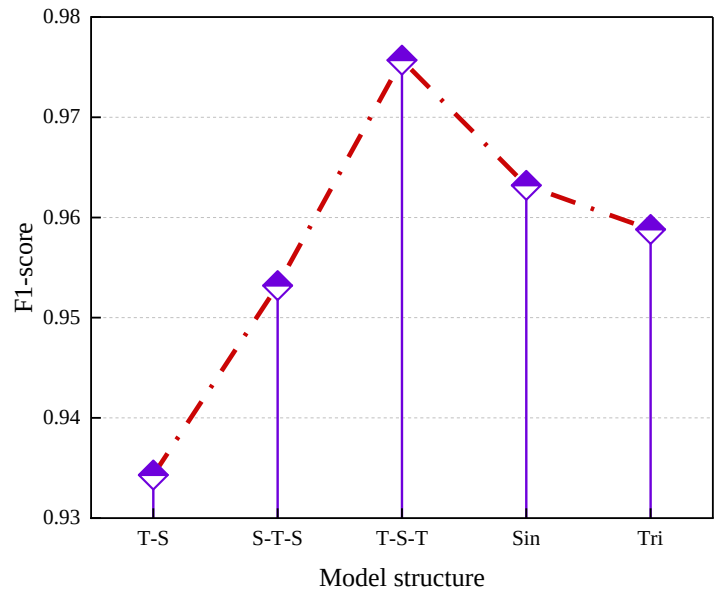

Fig. 19. Effect of model structure on localization performance

The results demonstrate that the proposed model structure in this paper achieves the highest F1-score value, indicating improved extraction of spatio-temporal features relevant to the attack. When the number of spatio-temporal graph convolution modules is small, feature extraction is insufficient. Conversely, when there are too many modules, issues like overfitting, computational complexity, and gradient vanishing can occur. Thus, the model structure in this paper strikes a suitable balance in enhancing attack localization performance.

### 5.5. Time cost of localization model

In the context of DDIA localization, both training time and localization time are important criteria for evaluating the efficiency of the model. The training time is indicative of the model's computational complexity, and the localization time reflects the timeliness of online data positioning. Fig. 20 illustrates the comparison of training times for different benchmark models across three test systems. Fig. 21, on the other hand, presents a comparison of localization times.

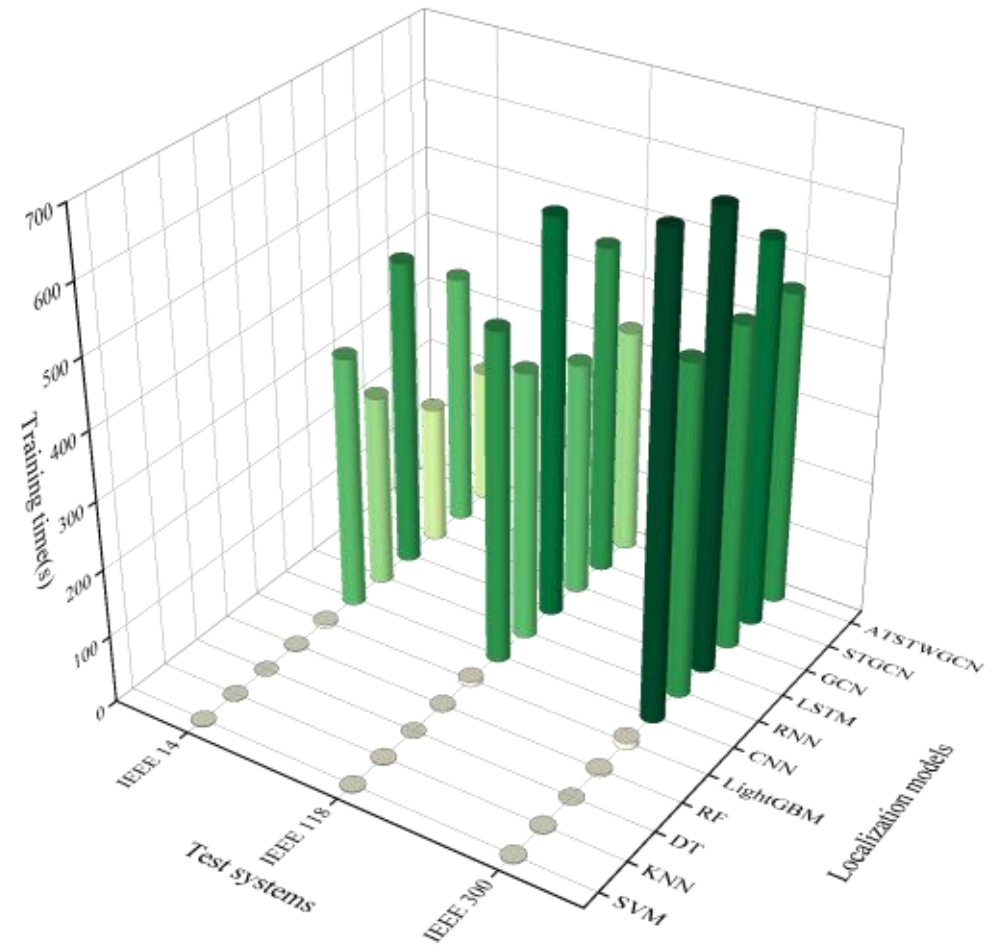

Fig. 20. Comparison of training time of different models for different test systems

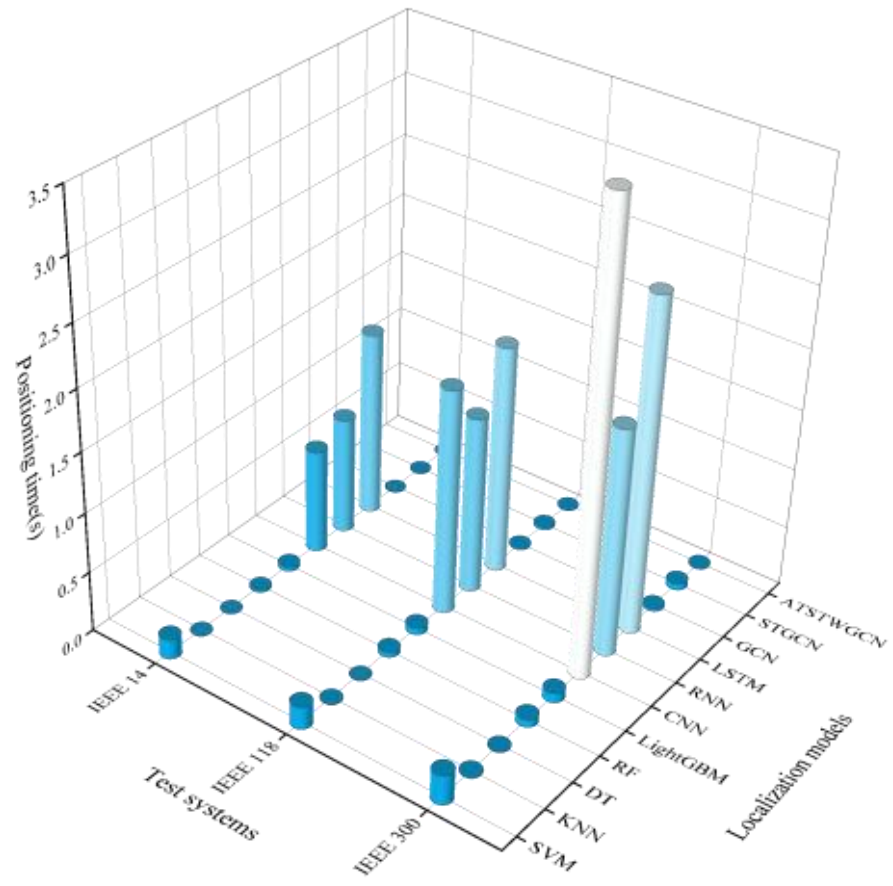

Fig. 21. Comparison of localization time of different models for different test systems

### 5.6. Noise-resistant Robustness of Localization Models

During the operation of a real power system, various environmental noise sources, including random disturbances, measurement errors, and communication errors, are inevitably introduced. To evaluate the robustness of the localization model against noise, different levels of noise intensity are introduced to the input dataset. This added noise is modeled to follow a normal distribution with a mean of 0 and a standard deviation ( $\varpi$ ), with the $\varpi$ range set between [0, 0.1]. Benchmark models that demonstrated better performance in prior studies are chosen for comparison, and the localization results for each model are depicted in Figure 22.

As the intensity of noise increases, the performance of the selected localization models deteriorates, making it more challenging to differentiate DDIAs from normal measurement data in high-noise environments. Notably, the CNN model experiences the swiftest decline in performance. This can be attributed to the spatial distance concealment feature of the DDIA, which poses difficulties for the CNN in accurately identifying DDIAs.

In contrast, the proposed model consistently outperforms the benchmark models across the entire range of noise intensity. When the standard noise deviation ( $\varpi$ ) is less than 0.04, the F1-score of the proposed model consistently remains above 0.97. Even at $\varpi$ = 0.10, the F1-score is 0.9333, which is also significantly better than the performance of other models. Furthermore, the proposed model exhibits the least fluctuation in performance, underscoring its strong robustness against noise and its stability.

This demonstrates that the proposed model effectively handles environmental noise, making it a robust and reliable choice for the power system DDIA localization.

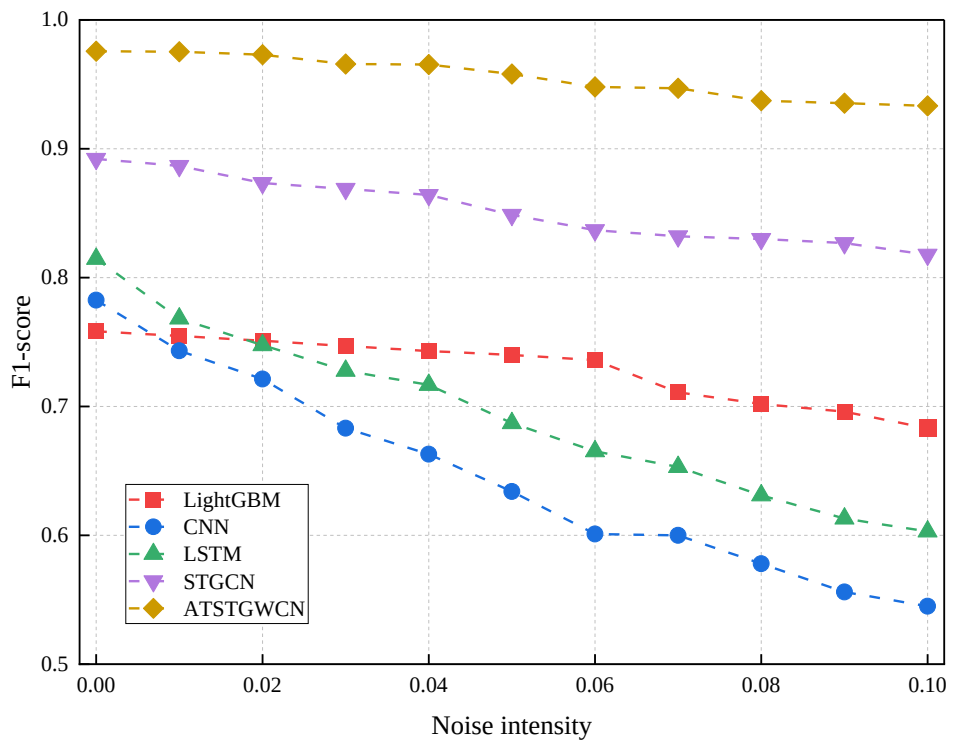

Fig. 22. Effect of noise intensity on model localization performance

## 6. CONCLUSION

For the novel form of DDIA in power systems, we have constructed a DDIA mathematical model that takes into account AC state estimation and incomplete topology. Additionally, we've proposed an attack localization method based on a spatio-temporal graph wavelet convolutional neural network, which introduces an attention mechanism. We have drawn the following conclusions based on experimental validation:

(1) By constructing a data injection attack that adheres to consistency, virtualization, destructiveness, and finiteness constraints, we have successfully bypassed bad data detection in state estimation. This renders distance-based attack detection methods ineffective and can have significant consequences for the power system, including the risk of line overload. The attacker's level of topology information significantly affects the success rate of the attack.

(2) In comparison to existing localization methods, this paper presents the advantage of a DDIA localization method based on the spatio-temporal attention mechanism, gated inflationary causal convolution, and graph wavelet convolution. This approach integrates time-sequential trend correlation and spatial topology correlation, as well as automates the extraction of spatio-temporal attack features. Through an extensive set of experiments, we have demonstrated that our proposed method effectively localizes the occurrence location of the DDIA. It outperforms other benchmark models and existing methods, enhancing accuracy, precision, recall, F1-score, and reducing the false alarm rate of attack localization while improving computational efficiency. The model's structure and scale factor can be adapted to DDIA localization tasks in various scenarios.

(3) The proposed ATSTWGCN exhibits strong generalization across different attack intensities, various test system topologies and sizes, diverse model parameters and structures, and robustness in noisy environments.

In future research, we aim to consider a broader range of attack types, design localization models capable of identifying multiple types of attacks, and implement more adaptive attack defense strategies.

## ACKNOWLEDGEMENTS

This work was supported by the National Natural Science Foundation of China (No. 52377081) and Natural Science Foundation of Jilin Province (20220101234JC).